\documentclass[12pt]{article}

\usepackage[margin=1in]{geometry}

\usepackage{enumitem} 

\usepackage{centernot}

\usepackage{times}

\usepackage{setspace} 
\usepackage{caption}
\usepackage{amsmath}
\usepackage{amssymb}
\usepackage{amsthm} 
\usepackage{mathtools} 

\usepackage[utf8]{inputenc} 
\usepackage[english]{babel}

\usepackage{graphicx} 

\usepackage{float}

\usepackage{siunitx} 
\newtheorem{theorem}{Theorem}
\newtheorem{assumption}{Assumption}

\newtheorem{condition}{Condition}

\usepackage[title]{appendix}

\numberwithin{equation}{section}

\newcommand\blfootnote[1]{
  \begingroup
  \renewcommand\thefootnote{}\footnote{#1}
  \addtocounter{footnote}{-1}
  \endgroup
}

\usepackage[hang]{footmisc}
\usepackage{hyperref}
\hypersetup{colorlinks=true, allcolors=black}

\usepackage[authoryear,round,sort,comma,longnamesfirst]{natbib} 
\usepackage{myapalike}
\begin{document}

\begin{titlepage}
\begin{center}
\linespread{1.2}
\Large{\textbf{Gaussian and Student's \textit{t} mixture vector autoregressive model with application to the  effects of the Euro area monetary policy shock}}\\

\vspace{0.5cm} 
\Large{Savi Virolainen}\\
\large{University of Helsinki}\\
\vspace{1.0cm}

\begin{abstract}
\noindent 
A new mixture vector autoregressive model based on Gaussian and Student's $t$ distributions is introduced. As its mixture components, our model incorporates conditionally homoskedastic linear Gaussian vector autoregressions and conditionally heteroskedastic linear Student's $t$ vector autoregressions. For a $p$th order model, the mixing weights depend on the full distribution of the preceding $p$ observations, which leads to attractive practical and theoretical properties such as ergodicity and full knowledge of the stationary distribution of $p+1$ consecutive observations. A structural version of the model with statistically identified shocks is also proposed. The empirical application studies the effects of the Euro area monetary policy shock. We fit a two-regime model to the data and find the effects, particularly on inflation, stronger in the regime that mainly prevails before the Financial crisis than in the regime that mainly dominates after it. The introduced methods are implemented in the accompanying R package gmvarkit.
\\[0.5cm]

\end{abstract}

\vfill

\blfootnote{The author thanks Markku Lanne, Mika Meitz, and Pentti Saikkonen for discussions and comments, which helped to improve this paper substantially. The author also thanks Leena Kalliovirta for the useful comments and the Research Council of Finland for the financial support (Grants 308628 and 347986).}
\blfootnote{Contact address: Savi Virolainen, Faculty of Social Sciences, University of Helsinki, P. O. Box 17, FI–00014 University of Helsinki, Finland; e-mail: savi.virolainen@helsinki.fi. ORCiD ID: 0000-0002-5075-6821.}

\end{center}
\end{titlepage}

\section{Introduction}
Nonlinear vector autoregressive (VAR) models are useful for modelling series in which the data generating dynamics vary in time. Such variation may arise due to wars, crises, business cycle fluctuations, or policy shifts, for example. Markov-switching vector autoregressive (MS-VAR) models \citep{Krolzig:1997} have been particularly popular (e.g., \citealp{Garcia+Schaller:2002}; \citealp{Peersman+Smets:2002}; \citealp{Lo+Piger:2005}; and \citealp{Dolado+MariaDolores:2006}) due to the capability to flexibly model series exhibiting discrete regime-switches. Also logistic smooth transition vector autoregressive (LSTVAR) models \citep{Anderson+Vahid:1998} have been widely employed  (e.g., \citealp{Weise:1999}; \citealp{Auerbach+Gorodnichenko:2012}; and \citealp{Caggiano+Castelnuovo+Colombo+Nodari:2015}), as they allow to capture gradual shifts in the dynamics of the data with clear interpretations of the regimes. 

Although useful, MS-VAR and LSTVAR models have limited switching dynamics: in the former, they depend only on the preceding regime, while in the latter, they depend solely on the levels of the switching variables. As a result, these models cannot capture more complex switching dynamics that may depend on a variety of statistical properties of the data. The Gaussian mixture vector autoregressive (GMVAR) model of \cite{Kalliovirta+Meitz+Saikkonen:2016}, in contrast, incorporates switching dynamics that, for a $p$th order model, depend on the full distribution of the previous $p$ observations. Specifically, the greater the relative weighted likelihood of a regime is, the more likely the process is to generate an observation from it. This more data-driven approach facilitates capturing complex switching dynamics that do not depend on the choice of switching variables, and it leads to attractive theoretical properties such as ergodicity and full knowledge of the stationary distribution of $p+1$ consecutive observations. On the other hand, it is not always clear how to interpret the regimes of the GMVAR model. Moreover, since the regimes of the GMVAR model are conditionally homoskedastic linear Gaussian VARs, its capability to capture strong conditional heteroskedasticity and high kurtosis are rather limited.

In order to address the limitations of the extant models, this paper introduces a new mixture vector autoregressive model that is closely related to the GMVAR model but allows the regimes to be conditionally heteroskedastic Student's $t$ VARs. This model, which we refer to as the Gaussian and Student's $t$ mixture vector autoregressive (G-StMVAR) model, thereby accommodates conditionally homoskedastic linear Gaussian VARs and conditionally heteroskedastic linear Student's $t$ VARs as its regimes (or mixture components). The G-StMVAR model can be described as the multivariate counterpart of the Gaussian and Student's $t$ mixture autoregressive (G-StMAR) model of \cite{Virolainen:2022}, and if all of its regimes are Student's $t$ VARs, the multivariate counterpart of the Student's $t$ mixture autoregressive (StMAR) model of \cite{Meitz+Preve+Saikkonen:2023} 
is obtained as a special case. At each time point, the G-StMVAR model generates an observation from one of its mixture components that is randomly selected according to the probabilities given by the mixing weights.

Both types of mixture components in the G-StMVAR model have the same form for the conditional mean, a linear function of the preceding $p$ observations, but their conditional covariance matrices are different. The linear Gaussian VARs have constant conditional covariance matrices, while those of the linear Student's $t$ VARs are products of a constant covariance matrix and a time-varying scalar that depends on the quadratic form of the previous $p$ observations. In contrast to the GMVAR model, our model is, hence, able to capture excess kurtosis and conditional heteroskedasticity also within the regimes. 

Our specification of the conditional covariance matrix is not as general as the conventional multivariate autoregressive conditional heteroskedasticity (ARCH) process that allows the entries of the conditional covariance matrix to vary relative to each other \citep[e.g., ][Section16.3]{Lutkepohl:2005}. It is, nonetheless, convenient for establishing stationarity properties similar to the linear Gaussian VARs. Our specification is also parsimonious, as it only depends on the covariance matrix, degrees of freedom, and autoregressive parameters. This parsimony is particularly advantegeous in mixture VARs, where the large number of parameters can often be a problem even without an ARCH component. 
 
In addition to the reduced form model, we propose a structural version of the G-StMVAR model with statistically identified shocks. Specifically, the shocks are identified (up to ordering and sign) by simultaneously orthogonalizing them in all regimes similarly to \cite{Virolainen:2024}, who applied the identification to a linear SVAR model with regime-switching volatility. Our identification method restricts the relative magnitudes of the impact responses of the variables to each shock time-invariant, but in contrast to \cite{Virolainen:2024}, the impulse responses are allowed to vary across the regimes after the impact period. Consequently, our model accommodates state-dependent impulse responses, which may additionally vary depending on the sign and size of the shock. Our model also has the interesting property that it allows to assess the likelihood of future regime switches due to each shock. 

Compared to conventional recursive identification, our method is particularly advantageous when the shock of interest would be ordered last or nearly last, which is typical in monetary policy shock applications involving nonlinear SVARs, for example (e.g., \citealp{Weise:1999}; \citealp{Peersman+Smets:2002}; \citealp{Garcia+Schaller:2002}; \citealp{Lo+Piger:2005}; \citealp{Hoppner+Melzer+Neumann:2008}; \citealp{Pellegrino:2018}; and \citealp{Burgard+Neuenkirch+Nockel:2019}). This is because when the shock of interest (and the corresponding variable) is ordered last, recursive identification restricts the impact responses of the other variables to zero and thus time-invariant, while our methods allows them to deviate from zero. In line with the statistical identification literature, labelling the identified shocks requires external information, for example, in the form of economic short-run restrictions. Due to the statistical identification, the additional economic restrictions are also testable. 

Mixture VARs have been previously applied in structural analysis by \cite{Kalliovirta+Malinen:2020}, who identify the shocks in the GMVAR model by restricting the reduced form error covariance matrices and allow the impact responses of the variables to vary relative to each other across the regimes. They estimate the impulse response functions for each regime separately as if each of them was a linear VAR. In contrast, we allow the regime to switch as a result of a shock and compute the true (generalized) impulse response functions \citep{Koop+Pesaran+Potter:1996} of the nonlinear SVAR. \cite{Burgard+Neuenkirch+Nockel:2019}, in turn, propose a mixture VAR with logistic mixing weights and recursively identified shocks. As opposed to \cite{Burgard+Neuenkirch+Nockel:2019}, our identification scheme is more flexible in the sense that it does not require many (or necessarily any) zero restrictions on the impact effects of the shocks. Moreover, in our model, the regime-switching probabilities depend on the full distribution of the preceding $p$ observations instead of just on the levels of the switching-variables. 

To illustrate the use of our model, we study the effects of the Euro area monetary policy shock. We fit a G-StMVAR model with two Student's $t$ regimes to monthly data covering the period from January 1999 to December 2021 and consisting of a number of macroeconomic variables. We find that one regime (Regime 1) is characterized by a negative (but volatile) output gap, and it mainly prevails after the Financial crisis, whereas the other regime (Regime 2) is characterized by a positive output gap and mainly dominates before the Financial crisis. The effects of the monetary policy shock are stronger in Regime 2 than in Regime 1, but asymmetries with respect to the sign and size of the shock are weak. In both regimes, a contractionary monetary policy shock decreases output gap significantly and persistently. However, while the price level decreases significantly and permanently in Regime 2, the decrease is insignificant in Regime 1. 

The rest of this paper is organized as follows. Section~\ref{sec:linvar} introduces the linear Student's $t$ VAR and establishes its stationarity properties. Section~\ref{sec:stmvar} introduces the G-StMVAR model and discusses its properties. Section~\ref{sec:structural_model} introduces the structural G-StMVAR model, and Section~\ref{sec:girf} discusses impulse responses analysis. Section~\ref{sec:estimation} discusses estimation of the model parameters by the method of maximum likelihood (ML) and establishes the asymptotic properties of the ML estimator. Section~\ref{sec:building_gstmvar} discusses a strategy for building a G-StMVAR model, and Section~\ref{sec:empap} presents the empirical application. Appendix~\ref{sec:techdetlinvar} provides technical details related to Section~\ref{sec:linvar}, and in Appendix~\ref{ap:propt}, the density functions and some properties of the Gaussian and Student's $t$ distributions are discussed. Appendix~\ref{ap:proofs} gives proofs of the stated theorems, Appendix~\ref{sec:montecarlo} describes a Monte Carlo algorithm for estimating the generalized impulse response function, and Appendix~\ref{sec:empdetails} provides details on the empirical application. Finally, the introduced methods are implemented to the accompanying CRAN distributed R package gmvarkit \citep{gmvarkit}, which provides tools for estimation and other numerical analysis of the introduced model.

Throughout this paper, we use the following notation. We write $x=(x_1,...,x_n)$ for the column vector $x$ where the components $x_i$ may be either scalars or (column) vectors. The notation $x\sim n_d(\mu,\Sigma)$ signifies that the random vector $x$ has a $d$-dimensional Gaussian distribution with mean $\mu$ and (positive definite) covariance matrix $\Sigma$, and $n_d(\cdot;\mu,\Sigma)$ denotes the corresponding density function. Similarly, $x\sim t_d(\mu,\Sigma,\nu)$ signifies that $x$ has a $d$-dimensional $t$-distribution with mean $\mu$, (positive definite) covariance matrix $\Sigma$, and degrees of freedom $\nu$ (assumed to satisfy $\nu>2$), and $t_d(\cdot;\mu,\Sigma,\nu)$ denotes the corresponding density function. The $(d\times d)$ identity matrix is denoted by $I_d$, $\otimes$ denotes the Kronecker product, and $\boldsymbol{1}_d$ denotes a  $d$-dimensional vectors of ones.

\section{Linear Gaussian and Student's \textit{t} Vector Autoregressions}\label{sec:linvar}
The G-StMVAR model accommodates two types of mixture components: conditionally homoskedastic linear Gaussian vector autoregressions and conditionally heteroskedastic linear Student's $t$ vector autoregressions. This section defines these linear vector autoregressions and establishes their stationarity properties. Consider the $d$-dimensional linear VAR model defined as 
\begin{equation}\label{eq:linearvar}
z_t = \phi_{0} + \sum_{i=1}^pA_iz_{t-1} + \Omega_t^{1/2}\varepsilon_t,  
\end{equation}
where the error process $\varepsilon_t$ is independent and identically distributed (IID), $\Omega_t^{1/2}$ is a symmetric square root matrix of the positive definite $(d\times d)$ covariance matrix $\Omega_t$ for all $t$, and $\phi_0\in\mathbb{R}^d$ is an intercept parameter. The $(d \times d)$ autoregression matrices are assumed to satisfy $\boldsymbol{A}_p \equiv [A_1:...:A_p]\in\mathbb{S}^{d\times dp}$, where
\begin{equation}\label{eq:statreg}
\mathbb{S}^{d\times dp}= \lbrace [A_1:...:A_p]\in\mathbb{R}^{d\times dp}: \det(I_d - \sum_{i=1}^pA_iz^i)\neq 0 \ \text{for} \ |z|\leq 1 \rbrace
\end{equation}
defines the usual stability condition of a linear VAR. The linear Gaussian VAR is obtained from Equation~(\ref{eq:linearvar}) by assuming that $\varepsilon_t$ follows the $d$-dimensional standard normal distribution and that the conditional covariance matrix is a constant, $\Omega_t=\Omega$. We will first establish the stationarity properties of the linear Gaussian VAR, and by making use of the introduced notation, we then introduce the linear Student's $t$ VAR.

Under the stability condition, the linear Gaussian VAR is stationary, and the following properties are obtained. Denoting $\boldsymbol{z}_t=(z_t,...,z_{t-p+1})$ and $\boldsymbol{z}_t^{+}=(z_t,\boldsymbol{z}_{t-1})$, it is well known that the stationary solution to~(\ref{eq:linearvar}) satisfies
\begin{align}\label{eq:gausdist}
\begin{aligned}
\boldsymbol{z}_t & \sim n_{dp}(\boldsymbol{1}_p\otimes\mu,\Sigma_{p}) \\
\boldsymbol{z}^{+}_t & \sim n_{d(p+1)}(\boldsymbol{1}_{p+1}\otimes\mu,\Sigma_{p+1}) \\
z_t|\boldsymbol{z}_{t-1} & \sim 
 n_d(\phi_{0} + \boldsymbol{A}_p\boldsymbol{z}_{t-1}, \Omega),
 \end{aligned}
\end{align}
where the last line defines the conditional distribution of $z_t$ given $\boldsymbol{z}_{t-1}$. Formulas of the quantities $\mu,\Sigma_p,\Sigma_{p+1}$ are given in Appendix~\ref{sec:techdetlinvar}. 

We show in Appendix~\ref{sec:techdetlinvar} (Theorem~\ref{thm:tvar}) that there exists linear Student's $t$ VARs with stationarity properties analogous~(\ref{eq:gausdist}). Specifically, such Student's $t$ VARs are obtained by assuming that $\varepsilon_t$ follows the $d$-dimensional Student's $t$ distribution with mean zero, identity covariance matrix and $\nu+dp$ degrees of freedom and by defining conditional covariance matrix of $z_t$ as 
\begin{equation}\label{eq:Omegat_lin}
\Omega_{t} = \frac{\nu - 2 + (\boldsymbol{z}_{t-1} - \boldsymbol{1}_p\otimes\mu)'\Sigma_p^{-1}(\boldsymbol{z} - \boldsymbol{1}_p\otimes\mu)}{\nu - 2 + dp}\Omega.
\end{equation}
Using the same notation as with the linear Gaussian VAR, the stationary solution to~(\ref{eq:linearvar}) for the above-defined Student's $t$ VAR satisfies (Theorem~\ref{thm:tvar} in Appendix~\ref{sec:techdetlinvar})
\begin{align}\label{eq:studentdist}
\begin{aligned}
\boldsymbol{z}_t & \sim t_{dp}(\boldsymbol{1}_p\otimes\mu,\Sigma_{p},\nu) \\
\boldsymbol{z}^{+}_t & \sim t_{d(p+1)}(\boldsymbol{1}_{p+1}\otimes\mu,\Sigma_{p+1},\nu) \\
z_t|\boldsymbol{z}_{t-1} & \sim t_d(\phi_{0} + \boldsymbol{A}_p\boldsymbol{z}_{t-1}, \Omega_t, \nu + dp).
\end{aligned}
\end{align}

Our Student's $t$ VAR has a conditional mean identical to the Gaussian VAR, but unlike the Gaussian VAR, it is conditionally heteroskedastic. Specifically, the conditional  covariance matrix~(\ref{eq:Omegat_lin}) consists of a constant covariance matrix that is multiplied by a time-varying scalar that depends on the quadratic form of the preceding $p$ observations through the autoregressive parameters. In this sense, the model has a ‘VAR($p$)–ARCH($p$)’ representation, but the ARCH type conditional variance is not as general as in the conventional multivariate ARCH process \citep[e.g., ][Section 16.3]{Lutkepohl:2005} that allows the entries of the conditional covariance matrix to vary relative to each other. Our model is, however, more parsimonious than the conventional VAR-ARCH model, as the conditional covariance depends only on the degrees of freedom and autoregressive parameters (in addition to the parameters in the constant covariance matrix). Student's $t$ VARs similar to ours have previously appeared at least in \cite{Heracleous:2003} and \cite{Poudyal:2012}.

\section{The Gaussian and Student's \textit{t} Mixture Vector Autoregressive Model}\label{sec:stmvar}
The G-StMVAR model can be described as a collection of linear autoregressive models each of which is a linear Gaussian VAR or a linear Student's $t$ VAR defined in Section~\ref{sec:linvar}. At each time point, the process generates an observation from one of its mixture components that is randomly selected according to the probabilities given by the mixing weights. This definition is formalized in the following. 

Let $y_t$, $t=1,2,...$, be the real valued $d$-dimensional time series of interest, and let $\mathcal{F}_{t-1}$ denote $\sigma$-algebra generated by the random vectors $\lbrace y_s, s<t \rbrace$. In a G-StMVAR model with autoregressive order $p$ and $M$ mixture components (or regimes), the observations $y_t$ are assumed to be generated by 
\begin{align}
y_t = & \sum_{m=1}^Ms_{m,t}(\mu_{m,t} + \Omega_{m,t}^{1/2}\varepsilon_{m,t}), \label{eq:def} \\
\mu_{m,t} = & \phi_{m,0} + \sum_{i=1}^pA_{m,i}y_{t-i},\label{eq:mu_mt}
\end{align}
where the following conditions \citep[similar to Condition~1 in][]{Kalliovirta+Meitz+Saikkonen:2016} hold.
\begin{condition}\label{cond:def}
\ 
\begin{enumerate}[label=(\alph*)]
\item For $m=1,...,M_1\leq M$,  the random vectors $\varepsilon_{m,t}$ are IID $n_d(0, I_d)$ distributed,  and for $m=M_1+1,..., M$, they are IID $t_d(0, I_d,\nu_m + dp)$ distributed. For all $m$,  $\varepsilon_{m,t}$ are independent of $\mathcal{F}_{t-1}$.
\item For each $m=1,...,M$‚ $\phi_{m,0}\in\mathbb{R}^d$,  $\boldsymbol{A}_{m,p} \equiv [A_{m,1}:...:A_{m,p}]\in\mathbb{S}^{d\times dp}$ (the set $\mathbb{S}^{d\times dp}$ is defined in (\ref{eq:statreg})), and $\Omega_m$ is positive definite. For $m=1,...,M_1$, the conditional covariance matrices are constants, $\Omega_{m,t}=\Omega_m$. For $m=M_1+1,...,M$, the conditional covariance matrices $\Omega_{m,t}$ are as in~(\ref{eq:Omegat_lin}), except that $\boldsymbol{z}_{t-1}$ is replaced with $\boldsymbol{y}_{t-1}=(y_{t-1},...,y_{t-p})$ and the regime specific parameters $\phi_{m,0}$, $\boldsymbol{A}_{m,p}$, $\Omega_m$,  $\nu_m$ are used to define the quantities therein. For $m=M_1+1,...,M$, also $\nu_m>2$.\label{cond:cond1b}
\item The unobservable regime variables $s_{1,t},...,s_{M,t}$ are such that at each $t$, exactly one of them takes the value one and the others take the value zero according to the conditional probabilities expressed in terms of the ($\mathcal{F}_{t-1}$-measurable) mixing weights $\alpha_{m,t}\equiv \mathrm{P}(s_{m,t}=1|\mathcal{F}_{t-1})$ that satisfy $\sum_{m=1}^M\alpha_{m,t}=1$.
\item Conditionally on $\mathcal{F}_{t-1}$, $(s_{1,t},...,s_{M,t})$ and $\varepsilon_{m,t}$ are assumed independent.
\end{enumerate}
\end{condition}
The conditions $\nu_m>2$ in \ref{cond:cond1b} are made to ensure the existence of second moments.  This definition implies that the G-StMVAR model generates each observation from one of its mixture components, a linear Gaussian or Student's $t$ vector autoregression discussed in Section~\ref{sec:linvar}, and that the mixture component is selected randomly according to the probabilities given by the mixing weights $\alpha_{m,t}$.  

Without loss of generality, the first $M_1$ mixture components are assumed to be linear Gaussian VARs, and the last $M_2\equiv M - M_1$ mixture components are assumed to be linear Student's $t$ VARs. If all the component processes are Gaussian VARs ($M_1=M$), the G-StMVAR model reduces to the GMVAR model of \cite{Kalliovirta+Meitz+Saikkonen:2016}. If all the component processes are Student's $t$ VARs ($M_1=0$), we refer to the model as the StMVAR model.

Equations (\ref{eq:def}) and (\ref{eq:mu_mt}) and Condition~\ref{cond:def} lead to a model in which the conditional density function of $y_t$ conditional on its past, $\mathcal{F}_{t-1}$,  is given as
\begin{equation}\label{eq:conddist}
f(y_t|\mathcal{F}_{t-1}) = \sum_{m=1}^{M_1}\alpha_{m,t}n_d(y_t;\mu_{m,t},\Omega_{m}) +  \sum_{m=M_1+1}^M\alpha_{m,t}t_d(y_t;\mu_{m,t},\Omega_{m,t},\nu_m+dp).
\end{equation}
The conditional densities $n_d(y_t;\mu_{m,t},\Omega_{m,t})$ and $t_d(y_t;\mu_{m,t},\Omega_{m,t},\nu_m+dp)$ are obtained from (\ref{eq:gausdist}) and (\ref{eq:studentdist}), respectively. The explicit expressions of the density functions are given in Appendix~\ref{ap:propt}. To fully define the G-StMVAR model it is then left to specify the mixing weights $\alpha_{m,t}$.

Analogously to \cite{Kalliovirta+Meitz+Saikkonen:2015}, \cite{Kalliovirta+Meitz+Saikkonen:2016}, \cite{Meitz+Preve+Saikkonen:2023}, and \cite{Virolainen:2022, Virolainen:2024}, we define the mixing weights as weighted ratios of the stationary densities of the regimes corresponding to the previous $p$ observations. To formally specify the mixing weights, we first define the following function for notational convenience. Let
\begin{equation}\label{eq:d_mdp}
d_{m,dp}(\boldsymbol{y};\mathbf{1}_p\otimes\mu_m,\Sigma_{m,p},\nu_m)=
\left\{\begin{matrix*}[l]
 n_{dp}(\boldsymbol{y};\mathbf{1}_p\otimes\mu_m,\Sigma_{m,p}), & \text{when} \ m \leq M_1, \\
 t_{dp}(\boldsymbol{y};\mathbf{1}_p\otimes\mu_m,\Sigma_{m,p},\nu_m), & \text{when} \ m > M_1,
\end{matrix*}\right.
\end{equation}
where the $dp$-dimensional densities $n_{dp}(\boldsymbol{y};\mathbf{1}_p\otimes\mu_m,\Sigma_{m,p})$ and $t_{dp}(\boldsymbol{y};\mathbf{1}_p\otimes\mu_m,\Sigma_{m,p},\nu_m)$ correspond to the stationary distribution of the $m$th regime (given in Equation~(\ref{eq:gausdist}) for the Gaussian regimes and in Theorem~\ref{thm:tvar} for the Student's $t$ regimes). Denoting $\boldsymbol{y}_{t-1}=(y_{t-1},...,y_{t-p})$, the mixing weights of the G-StMVAR model are defined as
\begin{equation}\label{eq:alpha_mt}
\alpha_{m,t} = \frac{\alpha_md_{m,dp}(\boldsymbol{y}_{t-1};\mathbf{1}_p\otimes\mu_m,\Sigma_{m,p},\nu_m)}{\sum_{n=1}^M \alpha_n d_{n,dp}(\boldsymbol{y}_{t-1};\mathbf{1}_p\otimes\mu_n,\Sigma_{n,p},\nu_n)},
\end{equation}
where $\alpha_m\in (0,1)$, $m=1,...,M$, are mixing weights parameters assumed to satisfy $\sum_{m=1}^M\alpha_m = 1$, $\mu_m = (I_d - \sum_{i=1}^pA_{m,i})^{-1}\phi_{m,0}$, and covariance matrix $\Sigma_{m,p}$ is given in Equations~(\ref{eq:gausquantities}), (\ref{eq:gausmatrices1}), and (\ref{eq:gausmatrices2}) (in Appendix~\ref{sec:techdetlinvar}) but using the regime specific parameters to define the quantities therein.  

Because the mixing weights are weighted ratios of the stationary densities of the regimes corresponding to the previous $p$ observations, the greater the relative weighted likelihood of a given regime is, the more likely the process it to generate an observation from it. This is a convenient feature for forecasting, and it also facilitates associating statistical characteristics and economic interpretations to the regimes. Moreover, it turns out that this specific formulation of the mixing weights leads to attractive theoretical properties such as full knowledge of the stationary distribution of $p+1$ consecutive observations and ergodicity of the process. These properties are summarized in Theorem~\ref{thm:statdist} below.

Before stating the theorem, a few notational conventions are provided. We collect the parameters of a G-StMVAR model to the $((M(d + d^2p + d(d+1)/2 + 2) - M_1 - 1)\times 1)$ vector $\boldsymbol{\theta}=(\boldsymbol{\vartheta}_1,...,\boldsymbol{\vartheta}_M,\alpha_1,...,\alpha_{M-1},\boldsymbol{\nu})$, where $\boldsymbol{\vartheta}_m=(\phi_{m,0},vec(\boldsymbol{A}_{m,p}),vech(\Omega_m))$ and $\boldsymbol{\nu}=(\nu_{M_1+1},...,\nu_M)$. The last mixing weight parameter is obtained as $\alpha_M=1-\sum_{m=1}^{M-1} \alpha_m$. The G-StMVAR model with autoregressive order $p$, and $M_1$ Gaussian and $M_2$ Student's $t$ mixture components is referred to as the G-StMVAR($p,M_1,M_2$) model, whenever the order of the model needs to be emphasized. 

\begin{theorem}\label{thm:statdist}
Consider the G-StMVAR process $y_t$ generated by (\ref{eq:def}),  (\ref{eq:mu_mt}), and (\ref{eq:alpha_mt}) with Condition~\ref{cond:def} satisfied. Then, $\boldsymbol{y}_t=(y_t,...,y_{t-p+1})$ is a Markov chain on $\mathbb{R}^{dp}$ with stationary distribution characterized by the density
\begin{equation}
f(\boldsymbol{y};\boldsymbol{\theta}) = \sum_{m=1}^{M_1}\alpha_m n_{dp}(\boldsymbol{y};\boldsymbol{1}_p\otimes\mu_{m},\Sigma_{m,p}) + \sum_{m=M_1+1}^M\alpha_mt_{dp}(\boldsymbol{y};\boldsymbol{1}_p\otimes\mu_{m},\Sigma_{m,p},\nu_m).
\end{equation}
Moreover, $\boldsymbol{y}_t$ is ergodic.
\end{theorem}
The stationary distribution is a mixture of $M_1$ $dp$-dimensional Gaussian distributions and $M_2$ $dp$-dimensional $t$-distributions with constant mixing weights $\alpha_m$. The proof of Theorem~\ref{thm:statdist} in Appendix~\ref{sec:thmstatproof} shows that the marginal stationary distributions of $1,...,p+1$ consecutive observations are likewise mixtures of Gaussian and $t$-distributions. This gives the mixing weight parameters $\alpha_m$‚ $m=1,..,M$, interpretation as the unconditional probabilities of an observation being generated from the $m$th component process. The unconditional mean, covariance, and first $p$ autocovariances are hence easily obtained similarly to the GMVAR model of \cite{Kalliovirta+Meitz+Saikkonen:2016}. 

Our StMVAR model features ARCH type conditional heteroskedasticity in the sense that the conditional covariance matrix of each regime depends on the quadratic form of the preceding $p$ observations through the covariance matrix, autoregressive, and degrees of freedom parameters (see the discussion at the end of Section~\ref{sec:linvar}). This property arises from utilizing the multivariate Student's $t$-distribution as the stationary distribution of the regimes. The use of the $t$-distribution thereby allows for parsimonious modelling of series that display fat tails and conditional heteroskedasticity within the regimes. This is particularly advantageous in the context of mixture VARs, as the large number of parameters may often be problematic even without an ARCH component. Appropriate modelling of kurtosis and conditional heteroskedasticity is important, because they may affect the endogenously determined regime-switching probabilities. Ignoring these factors would leave out potentially important dynamics that may affect the results of an empirical investigation. If some of the regimes have a constant conditional covariance matrix and zero excess kurtosis, we allow them to be conditionally homoskedastic linear Gaussian VARs, which leads to the G-StMVAR model. 

\section{Structural G-StMVAR Model}\label{sec:structural_model}
To facilitate structural analysis, the structural shocks must be identified by orthogonalizing the reduced form errors so that they conditional covariance matrix is a diagonal matrix. Consider the G-StMVAR model defined by (\ref{eq:def}), (\ref{eq:mu_mt}), and (\ref{eq:alpha_mt}) with Condition \ref{cond:def} satisfied. We write the structural G-StMVAR model as 
\begin{align}
y_t &= \sum_{m=1}^Ms_{m,t}(\phi_{m,0}+\sum_{i=1}^pA_{m,i}y_{t-i}) + B_te_t,  \label{eq:def_sgstmvar1} \\
u_t &\equiv B_te_t = \sum_{m=1}^M s_{m,t}\Omega_{m,t}^{1/2}\varepsilon_{m,t}, \label{eq:def_sgstmvar2}
\end{align}
where $e_t=(e_{1t},...,e_{dt})$ $(d \times 1)$ is an orthogonal structural error. The definition~(\ref{eq:def_sgstmvar2}) of the structural error is similar to the definition~(3.1) and (3.2) in \cite{Virolainen:2024}, who studied a linear SVAR model incorporating switching in the volatility regime, but with shocks arriving from Student's $t$ regimes in addition to (or instead of) Gaussian ones.

For the Gaussian regimes ($m=1,...,M_1$)‚ $\Omega_{m,t}=\Omega_m$, whereas for the Student's $t$ regimes ($m=M_1+1,...,M$), $\Omega_{m,t}=\omega_{m,t}\Omega_m$, where 
\begin{equation}
\omega_{m,t} = \frac{\nu_m - 2 + (\boldsymbol{y}_{t-1} - \boldsymbol{1}_p\otimes\mu_m)'\Sigma_{m,p}^{-1}(\boldsymbol{y}_{t-1} - \boldsymbol{1}_p\otimes\mu_m)}{\nu_m - 2 + dp}.
\end{equation}
The invertible $(d\times d)$ impact matrix $B_t$, which governs the contemporaneous relationships of the shocks, is time-varying and a function of $y_{t-1},..., y_{t-p}$. Following \cite{Virolainen:2024}, we define the impact matrix so that it captures the conditional heteroskedasticity of the reduced form error, and thereby amplifies a constant-sized structural shock accordingly. Appropriate modelling of conditional heteroskedasticity in the impact matrix is of interest because the (generalized) impulse response functions may be asymmetric with respect to the size of the shock. 

We have $\Omega_{u,t}\equiv\text{Cov}(u_t|\mathcal{F}_{t-1})=\sum_{m=1}^{M_1}\alpha_{m,t}\Omega_m + \sum_{m=M_1+1}^{M}\alpha_{m,t}\omega_{m,t}\Omega_m$, while the conditional covariance matrix of the structural error $e_t=B_t^{-1}u_t$ (which are not IID but martingale differences and thereby uncorrelated) is obtained as
\begin{equation}\label{eq:covet}
\text{Cov}(e_t|\mathcal{F}_{t-1})=\sum_{m=1}^{M_1}\alpha_{m,t}B_t^{-1}\Omega_mB_t'^{-1} + \sum_{m=M_1+1}^{M}\alpha_{m,t}\omega_{m,t}B_t^{-1}\Omega_mB_t'^{-1}.
\end{equation}
The impact matrix $B_t$ should be chosen so that the right side of Equation~(\ref{eq:covet}) is a diagonal matrix. \cite{Virolainen:2024} shows that any such impact matrix that simultaneously diagonalizes $\Omega_1,...,\Omega_M$ (and thus also $\Omega_{1,t},...,\Omega_{M,t}$) has linearly independent eigenvectors of the matrix $\Omega_m\Omega_1^{-1}$ as its columns. Moreover, he shows that under the following assumption and a constant normalization of the structural error's conditional covariance matrix, say, $\Omega_{u,t}=I_d$, the impact matrix is unique up to ordering of its columns and changing all signs in a column.\footnote{\cite{Virolainen:2024} shows the uniqueness of the impact matrix for a linear SVAR model with shocks arriving from a mixture of Gaussian distributions, but the results directly apply to our structural G-StMVAR model as well.} 

\begin{assumption}\label{as:eigenvalues}
Consider $M$ positive definite $(d\times d)$ covariance matrices $\Omega_m$,  $m = 1,...,M$, and denote the strictly positive eigenvalues of the matrices $\Omega_m\Omega_1^{-1}$ as $\lambda_{mi}$, $i=1,...,d$, $m=2,...,M$. Suppose that for all $i\neq j\in\lbrace 1,...,d\rbrace$, there exists an $m\in\lbrace 2,...,M\rbrace$ such that $\lambda_{mi}\neq\lambda_{mj}$.
\end{assumption}

Following \cite{Virolainen:2024} (and \cite{Lanne+Lutkepohl:2010} and \cite{Lanne+Lutkepohl+Maciejowska:2010}), we utilize the following matrix decomposition that is convenient for specifying the impact matrix. We decompose the error term covariance matrices as:
\begin{equation}\label{eq:omega_decomp}
\Omega_1 = WW' \quad \text{and} \quad \Omega_m=W\Lambda_mW', \quad m=2,...,M,
\end{equation}
where the diagonal of $\Lambda_m=\text{diag}(\lambda_{m1},...,\lambda_{md})$, $\lambda_{mi}>0$ ($i=1,...,d$), contains the eigenvalues of the matrix $\Omega_m\Omega_1^{-1}$ and the columns of the nonsingular $W$ are the related eigenvectors (which are identical for all $m$ by construction). When $M=2$, decomposition~(\ref{eq:omega_decomp}) always exists \citep[Theorem~A9.9]{Muirhead:1982}, but for $M\geq 3$ its existence requires that the matrices $\Omega_m\Omega_1^{-1}$ share the common eigenvectors in $W$. If this is not the case, the impact matrix does not exist \citep[see][Section~3.1]{Virolainen:2024}; however, its existence is testable.

Similarly to \cite{Virolainen:2024}, any scalar multiples of the columns of $W$ comprise an appropriate impact matrix, but only specific scalar multiples comprise the locally unique impact matrix associated with a given normalization of the structural error's conditional covariance matrix. Direct calculation shows that the impact matrix associated with the normalization $\text{Cov}(e_t|\mathcal{F}_{t-1})=I_d$ is obtained as
\begin{equation}\label{eq:B-matrix}
B_t = W(\sum_{m=1}^{M_1}\alpha_{m,t}\Lambda_m + \sum_{m=M_1 + 1}^{M}\alpha_{m,t}\omega_{m,t}\Lambda_m)^{1/2},
\end{equation}
where $B_tB_t'=\Omega_{u,t}$. Since $B_t^{-1}\Omega_mB_t'^{-1}=\Lambda_m(\sum_{n=1}^{M_1}\alpha_{n,t}\Lambda_n + \sum_{n=M_1+1}^M\alpha_{n,t}\omega_{n,t}\Lambda_n)^{-1}$, the impact matrix~(\ref{eq:B-matrix}) simultaneously diagonalizes $\Omega_{1},...,\Omega_{M}$, and $\Omega_{u,t}$ (and thereby also $\Omega_{1,t},...,\Omega_{M,t}$) for each $t$ so that $\text{Cov}(e_t|\mathcal{F}_{t-1}) = I_d$.

Our model identifies the shocks up to ordering and sign under Assumption~\ref{as:eigenvalues}, and therefore, global identification of the shocks is obtained by fixing the signs and ordering of the columns of $B_t$. The ordering of the columns can be fixed by choosing an arbitrary ordering for the eigenvalues in the diagonals of $\Lambda_m$, $m=2,..,M$. The signs, in turn, can be normalized by placing a single strict sign constraint in each column of $B_t$. The identification is, however, a statistical one, and it does not reveal which column of the impact matrix is related to which shock, nor does it necessarily lead to structural shocks with economic interpretations. Moreover, as the structure of the impact matrix in~(\ref{eq:B-matrix}) shows, it follows from the simultaneous diagonalization of the covariance matrices that, for each shock, the relative impact responses of the variables are restricted to be constant over time. More specifically, this identifying restriction implies that the impact responses to the $i$th shock $e_{1t}$ are $w_i(\sum_{m=1}^{M_1}\alpha_{m,t}\lambda_{mi} + \sum_{m=M_1 + 1}^{M}\alpha_{m,t}\omega_{m,t}\lambda_{mi})^{1/2}e_{1t}$, where $w_i$ $(d\times 1)$ is the $i$th column of $W$.

Labelling the identified structural shocks by economic shocks requires external information, for instance, in the form of economically motivated overidentifying restrictions as in \cite{Lanne+Lutkepohl:2010} and \cite{Virolainen:2024}. Such restrictions may readily be satisfied by the unrestricted estimate of $B_t$ (as in \cite{Virolainen:2024}), and if not, the validity of the overidentifying economic restrictions can be tested for (as in \cite{Lanne+Lutkepohl:2010}). However, the validity of Assumption~\ref{as:eigenvalues} required for full identification of the shocks is difficult to justify formally. If Assumption~\ref{as:eigenvalues} fails, further restrictions such as (untestable) zero restrictions on $B_t$ are required for the identification \citep[see the discussion in][Section~3.2]{Virolainen:2024}. 

\section{Impulse response analysis}\label{sec:girf}
The expected effects of the structural shocks in the the structural G-StMVAR model generally depend on the initial values, as well as on the sign and size of the shock. Additionally, the regime may also switch as a result of a shock. This makes the conventional way of calculating impulse responses unsuitable \citep[see, e.g.,][Chapter~4]{Kilian+Lutkepohl:2017}. However, the aforementioned features of our SVAR model can be captured with the generalized impulse response function (GIRF) \citep{Koop+Pesaran+Potter:1996}, defined as:
\begin{equation}\label{eq:girf}
\text{GIRF}(h,\delta_i,\mathcal{F}_{t-1}) = \text{E}[y_{t+h}|e_{it}=\delta,\mathcal{F}_{t-1}] - \text{E}[y_{t+h}|\mathcal{F}_{t-1}],
\end{equation}
where $h$ is the horizon, $e_{it}$ is $j$th element of $e_t$, and $\mathcal{F}_{t-1}=\sigma\lbrace y_{t-j},j>0\rbrace$ as before. The first term on the right side of (\ref{eq:girf}) is the expected realization of the process at time $t+h$ conditionally on a structural shock of sign and size $\delta_i \in\mathbb{R}$ in the $i$th element of $e_t$ at time $t$ and the previous observations. The latter term on the right side is the expected realization of the process conditionally on the previous observations only. The GIRF thus expresses the expected difference in the future outcomes when the structural shock of sign and size $\delta_i$ in the $i$th element of $e_t$ hits the system at time $t$ as opposed to all shocks being random. 

An interesting property of our SVAR model is that it also facilitates estimating the effects of shocks on the mixing weights. In other words, different from most nonlinear SVAR models, in our setup, it is possible to assess the likelihood of future regime switches due to each shock. The related impulse response functions are obtained by replacing $y_{t+h}$ by $\alpha_{m,t+h}$ in Equation~(\ref{eq:girf}). 

The G-StMVAR model has a $p$-step Markov property, so conditioning on (the $\sigma$-algebra generated by) the $p$ previous observations $\boldsymbol{y}_{t-1}=(y_{t-1},...,y_{t-p})$ is effectively the same as conditioning on $\mathcal{F}_{t-1}$ at time $t$ and later. The history $\boldsymbol{y}_{t-1}$ can be either fixed or random, but with a random history, the GIRF becomes a random vector. Using a fixed $\boldsymbol{y}_{t-1}$ is useful when the interest is in the effects of a shock at a particular point of time, whereas more general results are obtained by assuming that $\boldsymbol{y}_{t-1}$ follows some distribution. For example, the overall effects of a shock in the G-StMVAR model can be estimated by assuming that $\boldsymbol{y}_{t-1}$ follows the stationary distribution of the model (characterized in Theorem~\ref{thm:statdist}). On the other hand, if the interest is in the effects of the shock in particular state of the economy represented by a regime, $\boldsymbol{y}_{t-1}$ can be assumed to follow the stationary distribution of the corresponding regime (presented in Equations~(\ref{eq:gausdist}) and (\ref{eq:studentdist})). 

The GIRF and its distributional properties can be estimated with a Monte Carlo algorithm that generates (partial) realizations of the process and then takes the sample mean for a point estimate. If $\boldsymbol{y}_{t-1}$ is random and follows distribution $G$, the GIRF should be estimated for different values of $\boldsymbol{y}_{t-1}$ generated from $G$, and then the sample mean and sample quantiles can be taken to obtain the point estimate and confidence intervals that reflect the uncertainty about the initial value. Such an algorithm, adapted from \citet[pp. 135-136]{Koop+Pesaran+Potter:1996} and \citet[pp. 601-602]{Kilian+Lutkepohl:2017}, is given in Appendix~\ref{sec:montecarlo}.  

Our SVAR model defined in Section~\ref{sec:structural_model} assumes a single impact matrix that captures the conditional covariance matrix of the reduced form error, which varies according to the mixing weights. This specification allows the impact responses of the variables to vary in magnitude, but due to the simultaneous diagonalization of the error term covariance matrices, the impact response of any variable to each shock is restricted to be constant relative to the impact responses of the other variables (see the discussion in Section~\ref{sec:structural_model}). It would, of course, be possible to relax the restriction of constant relative impact effects, but doing so would require additional restrictions to be imposed to achieve identification. \cite{Bacchiocchi+Fanelli:2015}, and \cite{Angelini+Bacchiocchi+Caggiano+Fanelli:2019} have proposed such a model with exogenous change points of the volatility regime, where identification is based on combining heteroskedasticity with (untestable) zero restrictions on the impact matrices in different regimes. In addition to the untestable restrictions being potentially challenging to justify economically, their setup is not as straightforward to apply as ours.

Through regime switches, our SVAR model still accommodates impulse responses that vary in the periods after the impact depending on the initial state of the economy as well as on the sign and size of the shock. Specifically, as the regime may switch as a result of a shock, different signs and sizes of the shock may induce different responses of the mixing weights from different initial values $\boldsymbol{y}_{t-1}$, leading to nonlinear impulse responses in the periods after the impact period. This follows from the fact that the intercepts and AR matrices in~(\ref{eq:def})-(\ref{eq:mu_mt}) are not restricted to be constant across the regimes. While allowing for regime-specific autoregressive parameters is different from much of the previous statistical identification literature, it is always possible to test for the constancy of these parameters, and in case of non-rejection, proceed with a linear SVAR model with $M$ volatility regimes. This test and estimation of the linear SVAR model are also implemented in the accompanying R package 
gmvarkit \citep{gmvarkit}. For instance, in our empirical application in Section~\ref{sec:empap}, the null hypothesis of constancy is clearly rejected (see Appendix~\ref{sec:adequacy}).

\section{Estimation}\label{sec:estimation}
The parameters of the G-StMVAR model can be estimated by the method of maximum likelihood (ML), and the exact log-likelihood function is available, as we have established the stationary distribution of the process in Theorem~\ref{thm:statdist}. Suppose the observed time series is $y_{-p+1},...,y_0,y_1,...,y_T$ and that the initial values are stationary. Then, the log-likelihood function of the G-StMVAR model takes the form
\begin{equation}\label{eq:loglik1}
L(\boldsymbol{\theta})=\log\left(\sum_{m=1}^M\alpha_m d_{m,dp}(\boldsymbol{y}_0;\boldsymbol{1}_p\otimes\mu_m,\Sigma_{m,p},\nu_m) \right) + \sum_{m=1}^M l_t(\boldsymbol{\theta}),
\end{equation}
where $d_{m,dp}(\cdot;\boldsymbol{1}_p\otimes\mu_m,\Sigma_{m,p},\nu_m)$ is defined in (\ref{eq:d_mdp}) and
\begin{equation}\label{eq:loglik2}
l_t(\boldsymbol{\theta}) = \log\left(\sum_{m=1}^{M_1}  \alpha_{m,t}n_d(y_t;\mu_{m,t},\Omega_{m})  + \sum_{m=M_1 + 1}^M  \alpha_{m,t}t_d(y_t;\mu_{m,t},\Omega_{m,t},\nu_m + dp  ) \right).
\end{equation}
If stationarity of the initial values seems unreasonable, one can condition on the initial values and base the estimation on the conditional log-likelihood function, which is obtained by dropping the first term on the right side of (\ref{eq:loglik1}). The rest of this section assumes that estimation is based on the conditional log-likelihood function divided by the sample size, $L_T^{(c)}(\boldsymbol{\theta})=T^{-1}\sum_{m=1}^M l_t(\boldsymbol{\theta})$, i.e., the ML estimator $\hat{\boldsymbol{\theta}}_T$ maximizes $L_T^{(c)}(\boldsymbol{\theta})$.

If there are two regimes in the model ($M=2$), the structural G-StMVAR model discussed in Section~\ref{sec:structural_model} is obtained from the estimated reduced form model by decomposing the covariance matrices $\Omega_1,...,\Omega_M$ as in (\ref{eq:omega_decomp}). If $M\geq 3$ or overidentifying constraints are imposed on $B_t$ through $W$,  the model can be reparametrized with $W$ and $\Lambda_m$ ($m=2,...,M$) instead of $\Omega_1,...,\Omega_M$, and the log-likelihood function can be maximized subject to the new set of parameters and constraints. In this case, the decomposition (\ref{eq:omega_decomp}) is plugged into the log-likelihood function and $vech(\Omega_1),...,vech(\Omega_M)$ are replaced with $vec(W)$ and $\boldsymbol{\lambda}_2,...,\boldsymbol{\lambda}_M$ in the parameter vector $\boldsymbol{\theta}$,  where $\boldsymbol{\lambda}_m=(\lambda_{m1},...,\lambda_{md})$. Instead of constraining $vech(\Omega_1),...,vech(\Omega_M)$ so that $\Omega_1,...,\Omega_M$ are positive definite, we impose the constraints $\lambda_{mi}>0$ for all $m=2,...,M$ and $i=1,...,d$. 

Establishing the asymptotic properties of the ML estimator requires that it is uniquely identified. In order to achieve unique identification, the parameters need to be constrained so that the mixture components cannot be 'relabelled' to produce the same model with a different parameter vector. To that end, we assume that
\begin{align}\label{eq:identcond}
\begin{aligned}
& \alpha_1 > \cdots > \alpha_{M_1} > 0,  \ \alpha_{M_1+1} > \cdots > \alpha_M > 0, \ \text{and} \ \boldsymbol{\vartheta}_i = \boldsymbol{\vartheta}_j \ \text{only if any of the conditions} \\
& (1) \ 1\leq i = j \leq M, \ (2) \ i\leq M_1 < j, \ (3) \ i,j>M_1 \ \text{and} \ \nu_i\neq \nu_j, \ \text{is satisfied.}
\end{aligned}
\end{align}
In the case of the structural G-StMVAR model, identification also requires that Assumption~\ref{as:eigenvalues} is satisfied (see Section~\ref{sec:structural_model}).\footnote{With the appropriate zero restrictions on $W$, this condition can be relaxed, however \citep[see the related discussion in][Section~3]{Virolainen:2024}.} Then, identification of the structural model follows from the identification of the reduced form model.

We summarize the constraints imposed on the parameter space in the following assumption.
\begin{assumption}\label{as:mle}
The true parameter value $\boldsymbol{\theta}_0$ is an interior point of $\boldsymbol{\Theta}$,  which is a compact subset of $\lbrace \boldsymbol{\theta}=(\boldsymbol{\vartheta}_1,...,\boldsymbol{\vartheta}_M,\alpha_1,...,\alpha_{M-1},\boldsymbol{\nu})\in\mathbb{R}^{M(d + d^2p + d(d+1)/2)}\times (0,1)^{M-1}\times (2,\infty)^{M_2}:\boldsymbol{A}_{m,p}\in\mathbb{S}^{d\times dp},  \Omega_m$ is positive definite,  for all $m=1,...,M$, and (\ref{eq:identcond}) holds$\rbrace$.
\end{assumption}
Asymptotic properties of the ML estimator under the conventional high-level conditions are stated in the following theorem (which is proven in Appendix~\ref{sec:thmmleproof}).  Denote $\mathcal{I}(\boldsymbol{\theta})=E\left[\frac{\partial l_t(\boldsymbol{\theta})}{\partial \boldsymbol{\theta}}\frac{\partial l_t(\boldsymbol{\theta})}{\partial \boldsymbol{\theta}'} \right]$ and $\mathcal{J}(\boldsymbol{\theta})=E\left[\frac{\partial^2 l_t(\boldsymbol{\theta})}{\partial \boldsymbol{\theta}\partial\boldsymbol{\theta}'} \right]$.
\begin{theorem}\label{thm:mle}
Suppose that $y_t$ are generated by the stationary and ergodic G-StMVAR process of Theorem~\ref{thm:statdist} and that Assumption~\ref{as:mle} holds.  Then, $\hat{\boldsymbol{\theta}}_T$ is strongly consistent, i.e., $\hat{\boldsymbol{\theta}}_T\rightarrow \boldsymbol{\theta}_0$ almost surely. Suppose further that (i) $T^{1/2}\frac{\partial}{\partial\boldsymbol{\theta}_0}L_T^{(c)}(\boldsymbol{\theta}_0) \overset{d}{\rightarrow}N(0,\mathcal{I}(\boldsymbol{\theta}_0))$ with $\mathcal{I}(\boldsymbol{\theta}_0)$ finite and positive definite, (ii) $\mathcal{J}(\boldsymbol{\theta}_0)=-\mathcal{I}(\boldsymbol{\theta}_0)$, and (iii) $E[\sup_{\boldsymbol{\theta}\in\boldsymbol{\Theta}_0}|\frac{\partial^2 l_t(\boldsymbol{\theta})}{\partial\boldsymbol{\theta}\partial\boldsymbol{\theta}'}|]<\infty$ for some $\boldsymbol{\Theta}_0$, compact convex set contained in the interior of $\boldsymbol{\Theta}$ that has $\boldsymbol{\theta}_0$ as an interior point. Then  $T^{1/2}(\hat{\boldsymbol{\theta}}_T - \boldsymbol{\theta}_0)\overset{d}{\rightarrow} N(0,-\mathcal{J}(\boldsymbol{\theta}_0)^{-1})$.
\end{theorem}
Given consistency, conditions (i)-(iii) of Theorem~\ref{thm:mle} are standard for establishing asymptotic normality of the ML estimator, but their verification can be tedious. If one is willing to assume the validity of these conditions, the ML estimator has the conventional limiting distribution, implying that the approximate standard errors for the estimates are obtained as usual. Furthermore, the standard likelihood based tests are applicable as long as the number of mixture components is correctly specified.\footnote{This condition is important, because if the number of Gaussian or Student's $t$ type mixture components is chosen too large, some of the parameters are not identified causing the result of Theorem~\ref{thm:mle} to break down. This particularly happens when one tests for the number of regimes, as under the null some of the regimes are removed from the model. \cite{Meitz+Saikkonen:2021} have, however, recently developed such tests for mixture autoregressive models with Gaussian conditional densities. Developing a test for the number of a regimes in the G-StMVAR model is a major task and beyond the scope of this paper. Likewise, when testing whether a regime is a Gaussian VAR against the alternative that it is a Student's $t$ VAR, under the null, $\nu_m=\infty$ for the Student's $t$ regime $m$ to be tested, which violates Assumption~\ref{as:mle}.}

Finding the ML estimate amounts to maximizing the log-likelihood function defined in (\ref{eq:loglik1}) and (\ref{eq:loglik2}) over a high dimensional parameter space satisfying the constraints in Assumption~\ref{as:mle}. Due to the complexity of the log-likelihood function, numerical optimization methods are required. The maximization problem can be challenging in practice due to complex dependence of the mixing weights on the preceding observations, which induces a large number of modes to the surface of the log-likelihood function, and large areas to the parameter space, where it is flat in multiple directions. Also, the popular EM algorithm \citep{Redner+Walker:1984} is virtually useless here, as at each maximization step one faces a new optimization problem which not much simpler than the original one. Following \cite{Meitz+Preve+Saikkonen:2023} and \cite{Virolainen:2022, Virolainen:2024}, we therefore employ a two-phase estimation procedure in which a genetic algorithm is used to find starting values for a gradient based variable metric algorithm. The accompanying R package gmvarkit \citep{gmvarkit} employs a modified genetic algorithm that works similarly to the one described in \cite{Virolainen:2022}.

\section{Building a G-StMVAR Model}\label{sec:building_gstmvar}
Building a G-StMVAR model amounts to finding a suitable autoregressive order $p$, the number of Gaussian regimes $M_1$, and the number of Student's $t$ regimes $M_2$. Following the model selection strategy of \cite{Virolainen:2022} concerning the univariate counterpart of the G-StMVAR model, we propose taking advantage of the observation that the G-StMVAR model is a limiting case of the StMVAR model (in which all the mixture components are linear Student's $t$ VARs).  

It is easy to check that the linear Gaussian vector autoregression defined in Section~\ref{sec:linvar} is a limiting case of the linear Student's $t$ vector autoregression when the degrees of freedom parameter tends to infinity. As the mixing weights (\ref{eq:alpha_mt}) are weighted ratios of the stationary densities of the regimes, it then follows that a G-StMVAR($p,M_1,M_2$) model is obtained as a limiting case of the StMVAR($p,M$) model (or equivalently the G-StMVAR($p,0,M$) model) with the degrees of freedom parameters of the first $M_1$ regimes tending to infinity. Since a StMVAR($p,M$) model that is fitted to data generated by a G-StMVAR($p,M_1,M_2$) process is, therefore, asymptotically expected to get large estimates for the degrees of freedom parameters of the $M_1$ Gaussian regimes, we propose selecting the model by finding a suitable StMVAR model. If the fitted StMVAR model contains overly large estimates of the degrees of freedom parameters, the corresponding regimes should be switched to Gaussian VARs by estimating the appropriate G-StMVAR model.

For a strategy to find a suitable StMVAR model, we follow \cite{Kalliovirta+Meitz+Saikkonen:2015}, and suggest first considering the linear version of the model, that is, a StMVAR model with one mixture component. Partial autocorrelation functions, information criteria, and quantile residual diagnostics \citep[see][]{Kalliovirta+Saikkonen:2010} can be made use of for selecting the appropriate autoregressive order $p$. If the linear model is found inadequate, models with multiple mixture components can be examined. One should, however, be conservative with the choice of $M$, because if the number of regimes is chosen too large, some of the parameters are not identified. Adding new regimes to the model also vastly increases the number of parameters, and moreover, due to the increased complexity, it might be difficult to obtain the ML estimate if there are many regimes in the model. 

Overly large degrees of freedom parameters are redundant in the model, but their weak identification also causes numerical problems. Specifically, they induce a numerically nearly singular Hessian matrix of the log-likelihood function when evaluated at the estimate, which makes the approximate standard errors and the quantile residual diagnostic tests of \cite{Kalliovirta+Saikkonen:2010} often unavailable. Switching to the appropriate G-StMVAR model has little effect of on the fit of the model in the case of overly large estimates of degrees of freedom parameters, so it is then advisable. 

\section{Empirical Application}\label{sec:empap}
We employ our SVAR model to study the effects of the Euro area monetary policy shock. Its asymmetric effects of have been studied, among others, by \cite{Peersman+Smets:2002} and \cite{Dolado+MariaDolores:2006}, who found that it has larger effects on production during recessions than expansions. \cite{Pellegrino:2018} found the real effects of the monetary policy shock weaker during uncertain times than tranquil times, whereas \cite{Burgard+Neuenkirch+Nockel:2019} found the effects of contractionary monetary policy shocks stronger but less enduring during "crisis" than during "normal times".

We consider monthly Euro area data covering the period from January 1999 to December 2021 ($276$ observations) and consisting of four variables: industrial production index (IPI), harmonized consumer price index (HCPI), Brent crude oil price (Europe, OIL), and an interest rate variable (RATE). Our policy variable is the interest rate variable, which is the Euro overnight index average (EONIA) from January 1999 to October 2008 and the \cite{Wu+Xia:2016} shadow rate from November 2008 to December 2021. The \cite{Wu+Xia:2016} shadow rate is not bounded by the zero-lower-bound and also quantifies unconventional monetary policy measures.\footnote{The IPI, HCPI, and EONIA were obtained from the European Central Bank Statistical Data Warehouse; the Brent crude oil prices were retrieved from the Federal Reserve Bank of St. Louis database; and the \cite{Wu+Xia:2016} shadow rate was obtained from the first author's website. The dataset used is included the CRAN distributed R package gmvarkit \citep{gmvarkit}.} 

\begin{figure}[!t]
    \centerline{\includegraphics[width=\textwidth - 2cm]{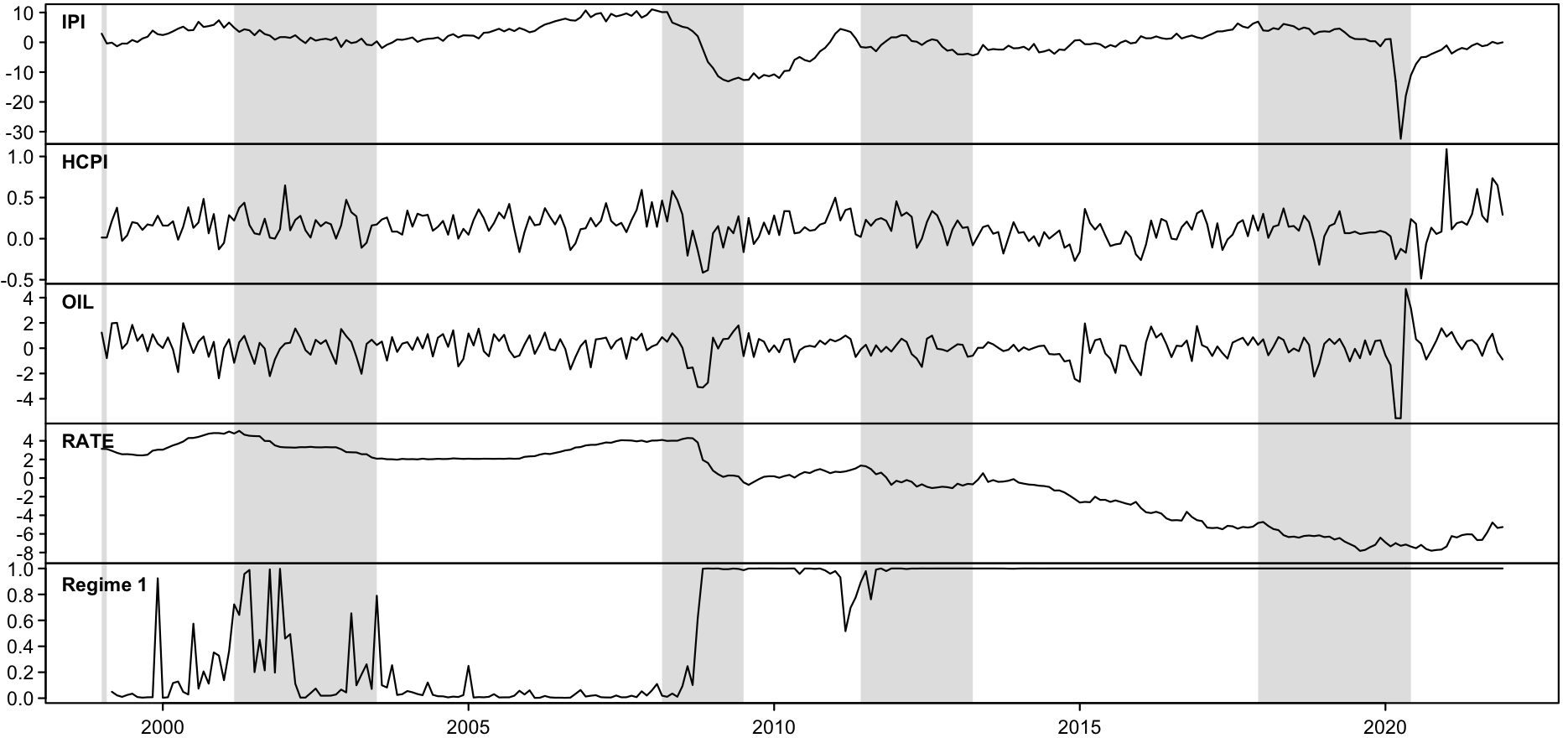}}
    \caption{Monthly Euro area series covering the period from January 1999 to December 2021. The top panel presents the cyclical component of the industrial production index (IPI) separated from the trend using the linear projection filter proposed by \cite{Hamilton:2018}. The second and third panels present the log-differences of the harmonized consumer price index (HCPI) and Brent crude oil prices (Europe, OIL) multiplied by hundred and ten, respectively. The fourth panel presents an interest rate variable, which is the EONIA from January 1999 to October 2008 and the \cite{Wu+Xia:2016} shadow rate from November 2008 to December 2021. The bottom panel shows the estimated mixing weights of Regime 1 of the fitted StMVAR($2,2$) model (the mixing weights of Regime 2 are unity minus the mixing weights of Regime 1). The shaded areas indicate the periods of Euro area recessions defined by the OECD.}
\label{fig:seriesplot}
\end{figure}

The IPI is detrended by first separating its cyclical component from the trend with the linear projection filter proposed by \cite{Hamilton:2018} and then considering the cyclical component.\footnote{Denoting the univariate, non-stationary time series as $y_t$, the filter defines its transient component at the time $t+h$ ($h>0$) as the ordinary least squares residuals from regressing $y_{t+h}$ on a constant and $y_t,...,y_{t-s+1}$. When $s$ is chosen larger than the order of integration, the residual process is stationary. We used the parameter values $h=24$ and $s=12$, as suggested by \cite{Hamilton:2018} for monthly data.}
It is thereby implicitly assumed that the monetary policy shock does not have permanent effects on real industrial production. Hereafter, we refer to the IPI's deviation from the trend as the output gap. The logs of HCPI and the oil price are detrended by taking first differences, whereas the interest rate variable is assumed stationary. For numerical reasons, we multiply the cyclical component of the IPI and the log-difference of HCPI by $100$, and the log-difference of OIL by $10$. The series are presented in Figure~\ref{fig:seriesplot}, where the shaded areas indicate the periods of Euro area recessions defined by the OECD \citep{OECD:2022}.

For selecting the order of our G-StMVAR model, we start by estimating one-regime StMVAR models with autoregressive orders $p=1,...,12$ and find that AIC is minimized at $p=1$. Next, we estimate a two-regime StMVAR model with $p=1$ but find this model somewhat inadequate. Consequently, we increase the autoregressive order to $p=2$, which increases the AIC but improves the model's adequacy. The overall adequacy of the StMVAR($2, 2$) model, i.e., G-StMVAR($p=2$, $M_1=0$, $M_2=2$) model, is found reasonable, so we employ it for the further analysis. Because the model does not contain large degrees of freedom parameter estimates, we do not consider incorporating Gaussian mixture components by switching to a G-StMVAR model with $M_1>0$. Details on model selection and the adequacy of the selected model are provided in Appendix~\ref{sec:adequacy}.

The estimated mixing weights of the StMVAR($2,2$) model are presented in the bottom panel of Figure~\ref{fig:seriesplot}. Regime 1 mainly prevails after the Financial crisis, but it obtains large mixing weights also before and during the early 2000's recession. Regime 2 dominates in the remaining time periods, that is, mainly before the Financial crisis. Since the prevailing regime starts switching sharply in October 2008, our model is consistent with the evidence that the ECB changed its reaction function after the bankruptcy of Lehman Brothers in September 2008 \citep{Gerlach+Lewis:2014}.\footnote{\cite{Gerlach+Lewis:2014} found that the ECB was cutting the interest rates faster at the time of the crisis, and that the ECB started a policy shift back in the late 2010. According our StMVAR model, however, the dominating regime never switches back to the pre-Financial crisis regime (in our sample period), although the second regime obtains mixing weights clearly larger than zero in the late 2010 and several relatively large mixing weights in the early 2011.}

Based on the unconditional means and marginal standard deviations of the regimes (presented in Table~\ref{tab:regimes} in Appendix~\ref{sec:characteristics}), the post-Financial crisis regime is characterized by negative but volatile output gap as well as lower inflation, oil price inflation, and interest rate interest rate than the pre-Financial crisis regime, which is characterized by positive output gap. The post-Financial crisis regime also exhibits higher kurtosis and overall volatility than the pre-Financial crisis regime (in all of the variables). Details on the characteristics of the regimes are discussed in Appendix~\ref{sec:characteristics}.

\subsection{Identification of the Monetary Policy Shock}\label{sec:identmone}
Decomposing the covariance matrices of the reduced form StMVAR($2,2$) model as in (\ref{eq:omega_decomp}) gives the following estimates for the structural parameters:
\begin{equation}\label{eq:estimates1}
\small
\hat{W}=\begin{bmatrix*}
\phantom{-}0.77 \ (0.512) & -0.94 \ (0.618) & \boldsymbol{1.82} \ (0.771) & -0.13 \ (0.271) \\
-0.12 \ (0.069) & \phantom{-}0.12 \ (0.077) & \boldsymbol{0.17} \ (0.079) & \phantom{-}0.03 \ (0.026) \\
\boldsymbol{-1.05} \ (0.447) & -0.39 \ (0.482) & 0.56 \ (0.306) & \phantom{-}0.21 \ (0.143) \\
\phantom{-}0.01 \ (0.016) & -0.02 \ (0.022) & 0.02 \ (0.030) & \boldsymbol{\phantom{-}0.50} \ (0.199) \\
\end{bmatrix*},
\ \
\hat{\lambda}_2 = \begin{bmatrix}
 0.50 \ (0.397) \\
 0.36 \ (0.291) \\
 0.19 \ (0.155) \\
 0.03 \ (0.021) \\
\end{bmatrix},
\end{equation}
where the ordering of the variables is $y_t=(\text{IPI}_t, \text{HCPI}_t, \text{OIL}_t, \text{RATE}_t)$, the estimates $\hat{\lambda}_{2i}$ are in decreasing order (which fixes an arbitrary ordering for the columns of $\hat{W}$), and approximate standard errors are given in parentheses next to the estimates. The estimates that deviate from zero by more than two times their approximate standard error are bolded. Many of the estimates $\hat{\lambda}_{2i}$ are somewhat close to each other, suggesting that it might not be clear whether Assumption~\ref{as:eigenvalues} is satisfied. The validity of Assumption~\ref{as:eigenvalues} is difficult to justify formally \citep[see][Section~3.2]{Virolainen:2024}; however, the robustness of our identification to its failure is discussed at the end of this section.  

The estimates and approximate standard errors in (\ref{eq:estimates1}) show that the fourth shock is the only shock that moves the interest variable significantly on impact, and it is also the only shock that moves production in the opposite direction. Therefore, we deem it as the monetary policy shock. The fourth shock, however, appears to move inflation and oil price inflation in the same direction as the interest rate variable, which is contrary to the standard economic theory that states that an increase in the nominal interest rate should decrease inflation by decreasing aggregate demand \cite[e.g.,][and the references therein]{Gali:2015}. 

Since the instantaneous increase of prices in response to a contractionary monetary policy shock is arguably implausible, we test for the zero restrictions that the instantaneous responses of inflation and oil price inflation are zero. The zero restrictions obtain the $p$-values $0.22$ and $0.15$ in a Wald test individually and the $p$-value $0.34$ jointly, so they are not rejected.\footnote{Interestingly, our results (suggesting that the zero restriction improves plausibility of the response of inflation) are contrary to \cite{Castelnuovo:2016}, who argued that muted response of inflation in the Euro area could be caused by misspecified zero restrictions in the impact matrix.} In addition to dampening the implausible impact responses of prices, the zero restrictions are useful because by Propositions~2 and 3 of \cite{Virolainen:2024}, they make the identification robust to the failure of Assumption~\ref{as:eigenvalues}, whose validity is difficult to justify formally. Specifically, the monetary policy shock is still identified if $\lambda_{2i}=\lambda_{2j}$ for any $i,j=1,2,3$ and additionally $\lambda_{2i}=\lambda_{24}$ for any one of $i=1,2,3$ (but the approximate standard errors and the Wald test results are valid only if $\lambda_{2i}$ are all distinct). Therefore, we proceed with the model involving these zero restrictions. 

\subsection{Generalized impulse response function}\label{sec:empap_girf}
We compute the GIRFs to the monetary policy shock to study its macroeconomic effects. As discussed in Section~\ref{sec:girf}, since we allow the regime to switch as a result of a shock, the structural StMVAR model accommodates asymmetries with respect the initial state of the economy as well as to the sign and size of the shock. Therefore, we compute the GIRFs conditionally on the initial values generated from the stationary distribution of each regime separately to positive (contractionary) and negative (expansionary) as well as to one-standard-error (small) and two-standard error (large) shocks. However, as the GIRFs to small and large shocks are similar, only the responses to small shocks are reported. We scale the GIRFs so that they correspond to a $25$ basis point instantaneous increase in the interest rate variable, making the responses to shocks of different signs and sizes comparable. 

\begin{figure}[!t]
    \centerline{\includegraphics[width=\textwidth - 2cm]{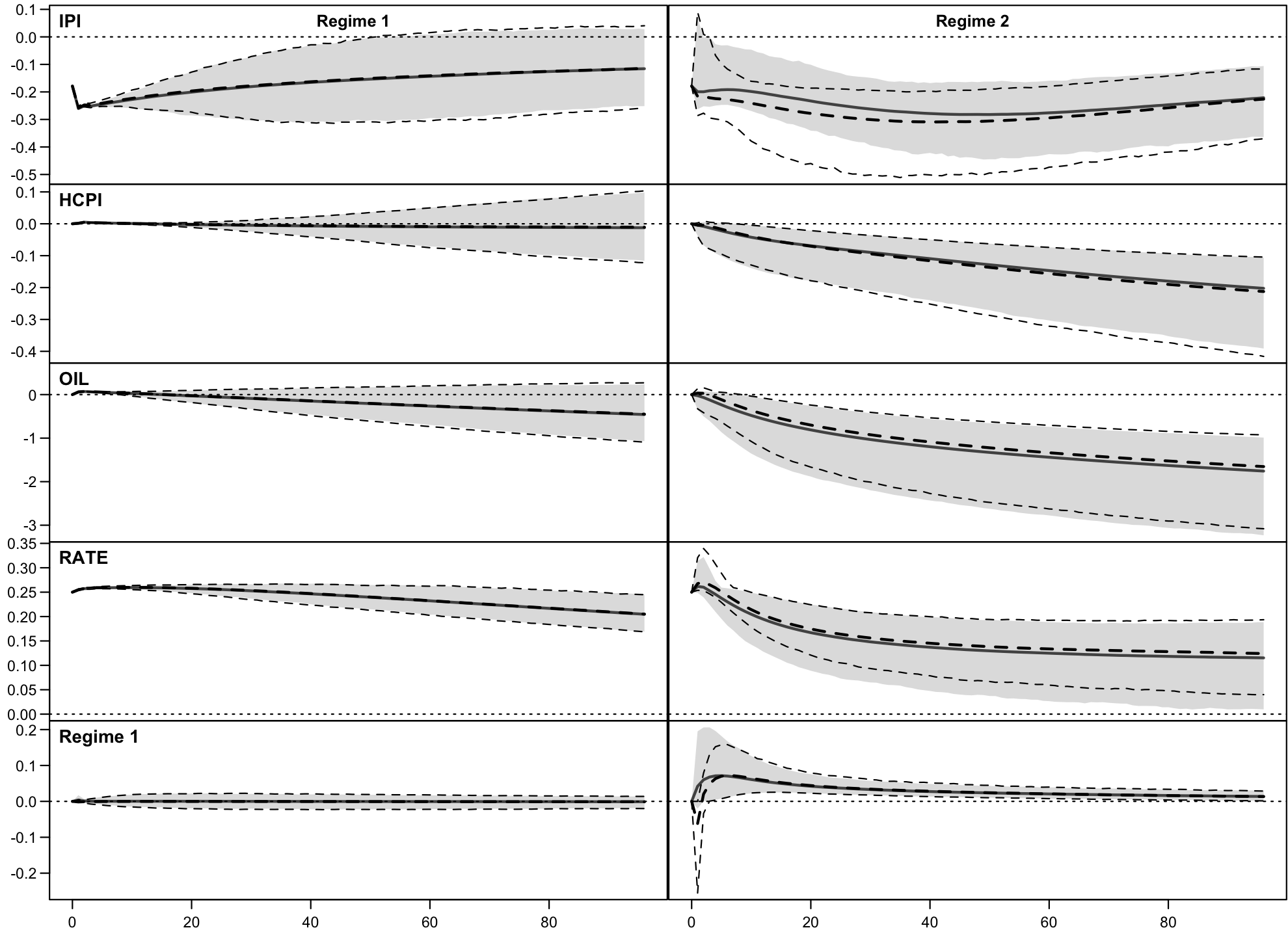}}
    \caption{Generalized impulse response functions $h=0,1,....,96$ months ahead to the monetary policy shock, based on $2500$ draws of initial values in the Monte Carlo algorithm presented in Appendix~\ref{sec:montecarlo}. From top to bottom, the responses of output gap, HCPI, oil price, the interest rate variable, and the mixing weights of Regime 1 are depicted in each row, respectively. The GIRFs of the HCPI and oil price are accumulated to (scaled) log-levels. The first column shows the responses to one-standard-error contractionary (grey solid line) and expansionary (black dashed line) shocks with the initial values generated from the stationary distribution of Regime 1. The second column shows the responses with the initial values generated from the Regime 2. The GIRFs are scaled to correspond $25$ basis point instantaneous increase of the interest rate variable. The $90\%$ confidence intervals that reflect the uncertainty about the initial value within the given regime are depicted are depicted with grey shaded areas for a contractionary shock and with black dashed lines for an expansionary shock.}
\label{fig:girfplot}
\end{figure}

Figure~\ref{fig:girfplot} presents the GIRFs $h=0,1,...,96$ months ahead computed for the identified monetary policy shock. The GIRFs to two-standard-error shocks are not reported for simplicity, since they seem similar to the GIRFs to one-standard-error shocks. In the post-Financial crisis Regime 1 (the left column in Figure~\ref{fig:girfplot}), the GIRFs to contractionary and expansionary shocks are practically identical. Output gap decreases strongly at impact in response to a contractionary monetary policy shock. On average, the peak response is in the first period, and then the response starts to slowly decay towards zero. The confidence intervals show that with some of the starting values the peak effect occurs later, however. The price level barely moves, but there is an insignificant permanent decrease. The oil price increases slightly for roughly fifteen months before it decreases persistently, and the response of the interest rate variable, in turn, is very persistent. The weak response of mixing weights of Regime 1 shows that the monetary policy shock has little effect on the regime switching probabilities. 

In the pre-Financial crisis Regime 2 (the right column in Figure~\ref{fig:girfplot}), the responses of the observable variables are quite symmetric with respect to the sign of the shock, but there are some differences in the responses of the mixing weights. Output gap decreases strongly at impact in response to a contractionary monetary policy shock. The response then decreases and peaks after several years before slowly decaying towards zero. The price level starts to steadily decrease after the impact period in response to a contractionary shock and keeps decreasing over our horizon of eight years. The oil price moves similarly to the consumer prices, whereas the response of the interest rate variable is very persistent. Interestingly, an expansionary shock significantly increases the probability of (the low-growth) Regime 1 in the first period after impact, but in the following periods, it significantly increases the probability of Regime 2. A contractionary shock, in turn, increases the probability of Regime 1 from the first period onwards. 

Overall, we find strong asymmetries with respect to the initial state of the economy, but the asymmetries seem weak with respect to the sign and size of the shock. The effects of the monetary policy shock on output gap and prices are stronger in the pre-Financial crisis Regime 2 than in the post-Financial crisis Regime 1. The real effects are significant in both of the regimes, but the inflationary effects are only minor in Regime 1, while they are substantial in Regime 2. Finally, the effects on the probabilities of future regime switches are clearly stronger in Regime 2.\footnote{For robustness, we compute the GIRFs also for the StMVAR($1,2$) suggested by AIC. We find the estimated mixing weights quite similar to the main specification: one of the regimes mainly prevails prior to the Financial crisis and one after it.  and the effects of the monetary policy shock, particularly on inflation, are stronger in the pre-Financial crisis regime. For comparison, we also consider GMVAR of \citep{Kalliovirta+Meitz+Saikkonen:2016} identified by heteroskedasticity as in \cite{Virolainen:2024} and the standard linear recursively identified Gaussian SVAR model. We find that the structural GMVAR model produces both price puzzles and output puzzles (the latter one meaning that the output gap increases in response a contractionary shock or decreases in response to an expansionary shock). The linear recursively identified Gaussian SVAR model (with $p=3$ based on AIC) performs even worse: in response to a contractionary shock, the output gap increases strongly and persistently, and prices rise permanently (the results are similar with the autoregressive order $p=1,2$).}

\section{Conclusion}
We introduced a new mixture vector autoregressive model, referred to as the G-StMVAR model, which has attractive theoretical and practical properties. The G-StMVAR model accommodates conditionally homoskedastic Gaussian VARs and conditionally heteroskedastic Student's $t$ VARs as its mixture components. The mixing weights are defined as weighted ratios of the stationary densities of the regimes corresponding to the previous $p$ observations. This specification is appealing, as it states that the greater the relative weighted likelihood of a regime is, the more likely the process is to generate an observation from it, which facilitates associating economic interpretations to the regimes. The specific formulation of the mixing weights also leads to attractive theoretical properties such as ergodicity and full knowledge of the stationary distribution of $p+1$ consecutive observations. Moreover, the maximum likelihood estimator of a stationary G-StMVAR model is strongly consistent, and therefore, it has the conventional limiting distribution under conventional high level conditions. 

The G-StMVAR model is a multivariate version of the G-StMAR model of \cite{Virolainen:2022}. As a special case, by assuming that all the mixture components are linear Student's $t$ VARs, we have also introduced a multivariate version of the StMAR model of \cite{Meitz+Preve+Saikkonen:2023}, which we call the StMVAR model. In addition to the reduced form model, we introduced a structural version of the G-StMVAR model with statistically identified shocks. The structural G-StMVAR model identifies the shocks by heteroskedasticity similarly to the SVAR model of \cite{Virolainen:2024}, but our model is nonlinear and accommodates asymmetries in the (generalized) impulse response functions. The introduced methods are implemented to the CRAN distributed R package gmvarkit \citep{gmvarkit} that provides a comprehensive set of tools for maximum likelihood estimation and other numerical analysis of the introduced models.

Our empirical application studied the effects of the Euro area monetary policy shock in a monthly data consisting of four macroeconomic variables. We fitted a two-regime StMVAR to the data and found that Regime 1 is characterized by a negative (but volatile) output gap, and it mainly prevails after the Financial crisis, whereas Regime 2 is characterized by a positive output gap and it mainly dominates before the Financial crisis. We found that the effects of the monetary policy shock are stronger in Regime 2 than in Regime 1. In both regimes, a contractionary monetary policy shock decreases the output gap significantly and persistently. However, while the price level decreases significantly and permanently in Regime 2, the decrease is insignificant in Regime 1. Asymmetries with respect to the sign and size of the shock were found weak in both regimes. 



\bibliography{masterrefs.bib}

\pagebreak

\begin{appendices}
\renewcommand{\thefigure}{\thesection.\arabic{figure}}
\renewcommand{\thetable}{\thesection.\arabic{table}}
\renewcommand{\thetheorem}{\thesection.\arabic{theorem}}
\setcounter{figure}{0}    
\setcounter{table}{0}
\setcounter{theorem}{0}

\section{Technical details related to Section~\ref{sec:linvar}}\label{sec:techdetlinvar}
This appendix provides formulas for the quantities related to the linear Gaussian and Student's $t$ VAR models. Denoting by $\Sigma(h)$, $h=0,\pm 1, \pm 2,...$, the lag $h$ autocovariance matrix of $z_t$, the quantities $\mu,\Sigma_p,\Sigma_1,\Sigma_{1p},\Sigma_{p+1}$ are given as \cite[see, e.g.,][pp. 23, 28-29]{Lutkepohl:2005}
\begin{align}\label{eq:gausquantities}
\begin{aligned}
\mu = & (I_d - \sum_{i=1}^pA_i)^{-1}\phi_0 & (d\times 1) \\
\text{vec}(\Sigma_p) = & (I_{(dp)^2} - \boldsymbol{A}\otimes\boldsymbol{A})^{-1}\text{vec}(\boldsymbol{\Omega}) & ((dp)^2\times 1)  \\
\Sigma_1 = & \Sigma(0) 
& (d\times d) \\
\Sigma(p) = & A_1\Sigma(p - 1) + \cdots + A_p\Sigma(0) & (d\times d) \\
\Sigma_{1p} = & [\Sigma(1):...:\Sigma(p-1):\Sigma(p)] = \boldsymbol{A}_p\Sigma_p & (d\times dp) \\
\Sigma_{p+1} = & 
\begin{bmatrix}
\Sigma_1       & \Sigma_{1p} \\
\Sigma_{1p}' & \Sigma_p
\end{bmatrix}
& (d(p+1) \times d(p+1))
\end{aligned}
\end{align}
where
\begin{equation}\label{eq:gausmatrices1}
\Sigma_p = 
\underset{(dp\times dp)}{\begin{bmatrix}
\Sigma(0) &  \Sigma(1) & \cdots & \Sigma(p-1) \\
\Sigma(-1) &  \Sigma(0) & \cdots & \Sigma(p-2) \\
\vdots        &   \vdots   & \ddots & \vdots  \\
\Sigma(-p+1) &  \Sigma(-p+2) & \cdots & \Sigma(0) \\
\end{bmatrix}},
\quad
\boldsymbol{\Omega} = 
\underset{(dp\times dp)}{\begin{bmatrix}
\Omega & 0 & \cdots & 0  \\
0            & 0 & \cdots & 0   \\
\vdots    &  \vdots & \ddots & \vdots \\
0            & 0 & \hdots &  0
\end{bmatrix}},
\end{equation}
and
\begin{equation}\label{eq:gausmatrices2}
\boldsymbol{A} = 
\underset{(dp\times dp)}{\begin{bmatrix}
A_1 & A_2 & \cdots & A_{p-1} & A_p \\
I_d  & 0     & \cdots & 0            & 0 \\
0     & I_d  &             & 0            & 0 \\
\vdots &   & \ddots & \vdots    & \vdots \\
0     & 0     & \hdots & I_d         & 0
\end{bmatrix}}.
\end{equation}

In order to construct a linear Student's $t$ VAR with stationary properties analogous to (\ref{eq:gausdist}), we consider the appropriate marginal distribution of $p+1$ consecutive observations. Then, a connection to the VAR in~(\ref{eq:linearvar}) is made through the conditional distribution, and finally this process and its stationary properties are formally established. Suppose that for a random vector in $\mathbb{R}^{d(p+1)}$ it holds that $(z,\boldsymbol{z})\sim t_{d(p+1)}(\boldsymbol{1}_{p+1}\otimes\mu,\Sigma_{p+1},\nu)$, where $\nu>2$. By the properties of a multivariate Student's $t$-distribution (given in Appendix~\ref{ap:propt}),
the conditional distribution of $z$ given $\boldsymbol{z}$ is $z|\boldsymbol{z}\sim t_d(\mu(\boldsymbol{z}),\Omega(\boldsymbol{z}),\nu + dp)$, where $\mu(\boldsymbol{z}) = \phi_{0} + \boldsymbol{A}_p\boldsymbol{z}$ and 
\begin{equation}\label{eq:convarlinear}
\Omega(\boldsymbol{z}) = \frac{\nu - 2 + (\boldsymbol{z} - \boldsymbol{1}_p\otimes\mu)'\Sigma_p^{-1}(\boldsymbol{z} - \boldsymbol{1}_p\otimes\mu)}{\nu - 2 + dp}\Omega.
\end{equation}

It is easy to see that a VAR of the form~(\ref{eq:linearvar}) that has the above-described conditional Student's $t$-distribution is obtained by assuming that $\varepsilon_t\sim t_d(0, I_d,\nu + dp)$ and that the conditional covariance matrix $\Omega_t$ is of the form~(\ref{eq:convarlinear}). The following theorem then formally establishes this Student's $t$ VAR and its stationary properties \citep[which is analogous to Theorem 1 in][considering a univariate version of the Student's $t$ autoregression]{Meitz+Preve+Saikkonen:2023}.
\begin{theorem}\label{thm:tvar}
Suppose $\phi_0\in\mathbb{R}^d$,  $[A_1:...:A_p]\in\mathbb{S}^{d\times dp}$,$\Omega\in\mathbb{R}^{d\times d}$ is positive definite, and that $\nu > 2$. Then,  there exists a process $\boldsymbol{z}_t = (z_t,...,z_{t-p+1})$ $(t=0,1,2,...)$ with the following properties.
\enumerate[label=(\roman*)]{
\item The process $\boldsymbol{z}_t$ is a Markov chain on $\mathbb{R}^{dp}$ with a stationary distribution characterized by the density function $t_{dp}(\boldsymbol{1}_{p}\otimes\mu,\Sigma_{p},\nu)$. When $\boldsymbol{z}_0\sim  t_{dp}(\boldsymbol{1}_p\otimes\mu,\Sigma_p,\nu)$,  we have,  for $t=1,2,...,$ that $\boldsymbol{z}_t^+\sim t_{d(p+1)}(\boldsymbol{1}_{p+1}\otimes\mu,\Sigma_{p+1},\nu)$ and the conditional distribution of $z_t$ given $\boldsymbol{z}_{t-1}$ is
\begin{equation}
z_t|\boldsymbol{z}_{t-1}\sim t_d(\mu(\boldsymbol{z}_{t-1}),\Omega(\boldsymbol{z}_{t-1}),\nu + dp).
\end{equation}\label{thm1:condist}
\item Furthermore, for $t=1,2,...$, the process $z_t$ has the representation
\begin{equation}\label{eq:vararch}
z_t = \phi_0 + \sum_{i=1}^pA_iz_{t-i} + \Omega_t^{1/2}\varepsilon_t,
\end{equation}
where $\Omega_t = \Omega(\boldsymbol{z}_{t-1})$ is the conditional covariance matrix (see (\ref{eq:convarlinear})), $\varepsilon_t \sim \text{IID}\ t_d(0, I_d,\nu + dp)$, and $\varepsilon_t$ are independent of $\lbrace z_{t-j},j>0\rbrace$ for all $t$.\label{thm1:tvar}
}
\end{theorem}
Theorem~\ref{thm:tvar} is proven in Appendix~\ref{sec:thmtvarproof}. Analogously to the univariate linear Student's autoregression discussed in \cite{Meitz+Preve+Saikkonen:2023}, the results \ref{thm1:condist} and \ref{thm1:tvar} in Theorem~\ref{thm:tvar} are comparable to properties (\ref{eq:gausdist}) and (\ref{eq:linearvar}) of the Gaussian counterpart. Part~\ref{thm1:condist} shows that both the stationary and conditional distributions of $z_t$ are $t$–distributions, whereas part~\ref{thm1:tvar} clarifies the connection to the standard VAR model. 

\section{Properties of Multivariate Gaussian and Student's \textit{t} Distributions}\label{ap:propt}
Denote a $d$-dimensional real valued vector by $y$. It is well known that the density function of a $d$-dimensional Gaussian distribution with mean $\mu$ and covariance matrix $\Sigma$ is
\begin{equation}
n_d(y;\mu,\Sigma) = (2\pi)^{-d/2}\text{det}(\Sigma)^{-1/2}\exp\left\lbrace -\frac{1}{2}(y -\mu)'\Sigma^{-1}(y - \mu) \right\rbrace .
\end{equation}
Similarly to \cite{Virolainen:2022} and \cite{Meitz+Preve+Saikkonen:2023}, we parametrize the Student's $t$-distribution using its covariance matrix as a parameter together with the mean and the degrees of freedom. The density function of such a $d$-dimensional $t$-distribution with mean $\mu$, covariance matrix $\Sigma$, and $\nu>2$ degrees of freedom is \citep[see, e.g., Appendix A in][]{Meitz+Preve+Saikkonen:2023}
\begin{equation}
t_d(y;\mu,\Sigma,\nu)=C_d(\nu)\text{det}(\Sigma)^{-1/2}\left(1+\frac{(y -\mu)'\Sigma^{-1}(y - \mu)}{\nu-2}\right)^{-(d+\nu)/2},
\end{equation}
where
\begin{equation}
C_d(\nu)=\frac{\Gamma\left(\frac{d+\nu}{2}\right)}{\sqrt{\pi^d(\nu-2)^d}\Gamma\left(\frac{\nu}{2}\right)},
\end{equation}
and $\Gamma\left(\cdot\right)$ is the gamma function.  We assume that the covariance matrix $\Sigma$ is positive definite for both distributions.

Consider a partition $X=(X_1,X_2)$ of either Gaussian or $t$-distributed (with $\nu$ degrees of freedom) random vector $X$ such that $X_1$ has dimension $(d_1\times1)$ and $X_2$ has dimension $(d_2\times1)$. Consider also a corresponding partition of the mean vector $\mu=(\mu_1,\mu_2)$ and the covariance matrix
\begin{equation}
\Sigma=
\begin{bmatrix}
\Sigma_{11} & \Sigma_{12} \\
\Sigma_{12}' & \Sigma_{22}
\end{bmatrix},
\end{equation}
where, for example, the dimension of $\Sigma_{11}$ is $(d_1\times d_1)$.  In the Gaussian case, $X_1$ then has the marginal distribution $n_{d_1}(\mu_1,\Sigma_{11})$ and $X_2$ has the marginal distribution $n_{d_2}(\mu_2,\Sigma_{22})$.  In the Student's $t$ case,  $X_1$ has the marginal distribution $t_{d_1}(\mu_1,\Sigma_{11},\nu)$ and $X_2$ has the marginal distribution $t_{d_2}(\mu_2,\Sigma_{22},\nu)$ \citep[see, e.g., ][also in what follows]{Ding:2016}.

When $X$ has Gaussian distribution,  the conditional distribution of the random vector $X_1$ given $X_2=x_2$ is 
\begin{equation}
X_1\mid(X_2=x_2)\sim n_{d_1}(\mu_{1\mid2}(x_2),\Sigma_{1\mid2}(x_2)),
\end{equation}
where
\begin{align}
\mu (x_2)\equiv \mu_{1\mid2}(x_2) &= \mu_1+\Sigma_{12}\Sigma_{22}^{-1}(x_2-\mu_2) \quad \text{and}\label{eq:mux_gaus} \\ 
\Omega \equiv \Sigma_{1\mid2}(x_2) &= \Sigma_{11}-\Sigma_{12}\Sigma_{22}^{-1}\Sigma_{12}'. \label{eq:omegax_gaus}
\end{align}

When $X$ has $t$-distribution, the conditional distribution of the random vector $X_1$ given $X_2=x_2$ is 
\begin{equation}
X_1\mid(X_2=x_2)\sim t_{d_1}(\mu_{1\mid2}(x_2),\Sigma_{1\mid2}(x_2),\nu+d_2),
\end{equation}
where
\begin{align}
\mu (x_2) = \mu_{1\mid2}(x_2) &= \mu_1+\Sigma_{12}\Sigma_{22}^{-1}(x_2-\mu_2) \quad \text{and}\label{eq:mux} \\ 
\Omega (x_2) \equiv \Sigma_{1\mid2}(x_2) &= \frac{\nu-2+(x_2-\mu_2)'\Sigma_{22}^{-1}(x_2-\mu_2)}{\nu-2+d_2}(\Sigma_{11}-\Sigma_{12}\Sigma_{22}^{-1}\Sigma_{12}'). \label{eq:omegax}
\end{align}
In particular, we have
\begin{align}\label{eq:td_decomp}
n_d(x;\mu,\Sigma) &= n_{d_1}(x_1;\mu_{1|2}(x_2),\Sigma_{1|2}(x_2))n_{d_2}(x_2;\mu_2,\Sigma_{22}) \quad \text{and}\\
t_d(x;\mu,\Sigma,\nu) &= t_{d_1}(x_1;\mu_{1|2}(x_2),\Sigma_{1|2}(x_2),\nu+d_2)t_{d_2}(x_2;\mu_2,\Sigma_{22},\nu).
\end{align}

\section{Proofs}\label{ap:proofs}

\subsection{Proof of Theorem~\ref{thm:tvar}}\label{sec:thmtvarproof}
Corresponding to $\phi_0\in\mathbb{R}^d$, $\boldsymbol{A}_p\in \mathbb{S}^{d\times dp}$,  $\Omega\in\mathbb{R}^{d\times d}$ positive definite, and $\nu > 2$, define the notation $\mu$, $\Sigma_p$, $\Sigma_1(h)$ $(h=0,1,...,p)$,  $\Sigma_{1p}$, and $\Sigma_{p+1}$ as in (\ref{eq:gausquantities})-(\ref{eq:gausmatrices2}). Note that by construction and the assumption $\boldsymbol{A}_p\in \mathbb{S}^{d\times dp}$, $\Sigma_p$ and $\Sigma_{p+1}$ are symmetric positive definite block Toeplitz matrices with the $(d\times d)$ blocks $\Sigma_1(h)$, $h=0,1,...,p$. Analogously to \cite{Meitz+Preve+Saikkonen:2023}, we prove Part~\ref{thm1:condist} by constructing a $dp$-dimensional Markov chain $\boldsymbol{z}_t=(z_t,...,z_{t-p+1})$ ($t=1,2,...$) with the desired properties. Then, we make use of the theory of Markov chains to establish its stationary distribution. To that end, an appropriate transition probability measure and an initial distribution needs to be specified. For the former, assume that the transition probability of $\boldsymbol{z}_t$ is determined by the density function $t_d(z_t;\mu(\boldsymbol{z}_{t-1}),\Omega(\boldsymbol{z}_{t-1}),\nu + dp)$, where $\mu(\boldsymbol{z}_{t-1})$ and $\Omega(\boldsymbol{z}_{t-1})$ are obtained from (\ref{eq:mux}) and (\ref{eq:omegax}),  respectively, by replacing $\boldsymbol{x}_{2}$ with $\boldsymbol{z}_{t-1}$.  Because the distribution of the current observation depends only on the previous one, $\boldsymbol{z}_t$ is a Markov chain on $\mathbb{R}^{dp}$.  

Suppose the initial value $\boldsymbol{z}_0$ follows the $t$-distribution $t_{dp}(\boldsymbol{1}_p\otimes\mu, \Sigma_p,\nu)$. The properties of $t$-distribution (given in Appendix~\ref{ap:propt}) then imply that if $\boldsymbol{z}_t^+=(z_t,\boldsymbol{z}_{t-1})$, the density function of $\boldsymbol{z}_1^+$ is given by
\begin{equation}\label{eq:dens_zplus}
t_{d(p+1)}(\boldsymbol{z}_1^+;\boldsymbol{1}_{p+1}\otimes\mu,\Sigma_{p+1},\nu)
=
t_d(z_1;\mu(\boldsymbol{z}_0),\Omega(\boldsymbol{z}_0),\nu + dp)t_{dp}(\boldsymbol{z}_0;\boldsymbol{1}_{p}\otimes\mu,\Sigma_p,\nu).
\end{equation}
Thus, $\boldsymbol{z}_1^+\sim t_{d(p+1)}(\boldsymbol{1}_{p+1}\otimes\mu,\Sigma_{p+1},\nu)$, and from the block Toeplitz structure of $\Sigma_{p+1}$ it follows that the marginal distribution of $\boldsymbol{z}_1$ is the same as that of $\boldsymbol{z}_0$, i.e., $\boldsymbol{z}_1\sim t_{dp}(\boldsymbol{1}_p\otimes\mu, \Sigma_p,\nu)$. Hence, as $\boldsymbol{z}_t$ is a Markov chain, it has a stationary distribution characterized by the density $t_{dp}(\boldsymbol{1}_p\otimes\mu, \Sigma_p,\nu)$ \citep[][pp. 230-231]{Meyn+Tweedie:2009}, completing the proof of Part~\ref{thm1:condist}.

Denote by $\mathcal{F}_{t-1}^z$ the $\sigma$-algebra generated by the random variables $\lbrace z_s, s<t \rbrace$. To prove Part~\ref{thm1:tvar}, note that due to the Markov property, $z_t|\mathcal{F}_{t-1}^z\sim t_d(\mu(\boldsymbol{z}_0),\Omega(\boldsymbol{z}_0),\nu + dp)$. Therefore, the conditional expectation and conditional variance of $z_t$ given $\mathcal{F}_{t-1}^z$ can be written as
\begin{align}
E[z_t|\mathcal{F}_{t-1}^z] = & E[z_t|\boldsymbol{z}_{t-1}] =  \mu + \Sigma_{1p}\Sigma_p^{-1}(\boldsymbol{z}_{t-1} - \boldsymbol{1}_p\otimes\mu) =  \phi_{0} + \boldsymbol{A}_p\boldsymbol{z}_{t-1}, \\
Var[z_t|\mathcal{F}_{t-1}^z] = & Var[z_t|\boldsymbol{z}_{t-1}] = \frac{\nu - 2 + (\boldsymbol{z}_{t-1} - \boldsymbol{1}_p\otimes\mu)'\Sigma_p^{-1}(\boldsymbol{z}_{t-1} - \boldsymbol{1}_p\otimes\mu)}{\nu - 2 + dp}\Omega,
\end{align}
where $\Omega = \Sigma_1 - \Sigma_{1p}\Sigma_p^{-1}\Sigma_{1p}'$. We denote this conditional variance by $\Omega_t\equiv \Omega(\boldsymbol{z}_{t-1})$, which is positive definite due to the assumptions $\nu >2$ and that $\Sigma_p$ and $\Omega$ are both positive definite. Define the $(d\times 1)$ random vectors $\varepsilon_t$ as
\begin{equation}
\varepsilon_t \equiv \Omega_t^{-1/2}(z_t - \phi_0 - \boldsymbol{A}_p\boldsymbol{z}_{t-1}),
\end{equation}
where $\Omega_t^{-1/2}$ is a symmetric square root matrix of $\Omega_t^{-1}$. Conditionally on $\mathcal{F}_{t-1}^z$, $\varepsilon_t$ now follow the $t_d(0,I_d,\nu + dp)$ distribution, and therefore the 'VAR($p$)-ARCH($p$)' representation~(\ref{eq:vararch}) is obtained. Because this conditional distribution does not depend on $\mathcal{F}_{t-1}^z$, it follows that the unconditional distribution of $\varepsilon_t$ is also $t_d(0,I_d,\nu + dp)$. Hence, $\varepsilon_t$ is independent of $\mathcal{F}_{t-1}^z$ (or of $\lbrace z_s, s<t \rbrace$), and as the random vectors $\lbrace \varepsilon_s, s<t \rbrace$ are functions of $\lbrace z_s, s<t \rbrace$, $\varepsilon_t$ is also independent of $\lbrace \varepsilon_s, s<t \rbrace$. Thus, the proof of Part~\ref{thm1:tvar} is completed by concluding that the random vectors $\varepsilon_t$ are IID $t_d(0,I_d,\nu + dp)$ distributed.$\blacksquare$

\subsection{Proof of Theorem~\ref{thm:statdist}}\label{sec:thmstatproof}
The G-StMVAR process $\boldsymbol{y}_t$ is clearly a Markov chain on $\mathbb{R}^{dp}$. Let $\boldsymbol{y}_0=(y_0,...,y_{-p+1})$ be random vector whose distribution is characterized by the density $f(\boldsymbol{y}_0;\boldsymbol{\theta})= \sum_{m=1}^{M_1}\alpha_m n_{dp}(\boldsymbol{y}_0;\boldsymbol{1}_p\otimes\mu_{m},\Sigma_{m,p}) + \sum_{m=M_1+1}^M\alpha_mt_{dp}(\boldsymbol{y}_0;\boldsymbol{1}_p\otimes\mu_{m},\Sigma_{m,p},\nu_m)$. According to (\ref{eq:gausdist}), (\ref{eq:studentdist}), (\ref{eq:def}), (\ref{eq:mu_mt}), and (\ref{eq:alpha_mt}), the conditional density of $y_1$ given $\boldsymbol{y}_0$ is
\begin{align}
f(y_1|\boldsymbol{y}_0;\boldsymbol{\theta})  =& \sum_{m=1}^{M_1}\frac{\alpha_m n_{dp}(\boldsymbol{y}_0;\boldsymbol{1}_p\otimes\mu_m,\Sigma_{m,p})}{f(\boldsymbol{y}_0;\boldsymbol{\theta})} n_d(y_1;\mu_{m,1}(\boldsymbol{y}_0),\Omega_{m,1})\nonumber \\
&+ \sum_{m=M_1+1}^M\frac{\alpha_m t_{dp}(\boldsymbol{y}_0;\boldsymbol{1}_p\otimes\mu_m,\Sigma_{m,p},\nu_m)}{f(\boldsymbol{y}_0;\boldsymbol{\theta})} t_d(y_1;\mu_{m,1}(\boldsymbol{y}_0),\Omega_{m,1}(\boldsymbol{y}_0),\nu_m+dp)\\
=&  \sum_{m=1}^{M_1}\frac{\alpha_m}{f(\boldsymbol{y}_0;\boldsymbol{\theta})} n_{d(p+1)}((y_t,\boldsymbol{y}_0);\boldsymbol{1}_{p+1}\otimes\mu_m,\Sigma_{m,p+1})\nonumber\\
&+\sum_{m=M_1+1}^M\frac{\alpha_m}{f(\boldsymbol{y}_0;\boldsymbol{\theta})} t_{d(p+1)}((y_t,\boldsymbol{y}_0);\boldsymbol{1}_{p+1}\otimes\mu_m,\Sigma_{m,p+1},\nu_m).
\end{align}
The random vector $(y_1, \boldsymbol{y}_0)$ therefore has the density 
\begin{align}
\begin{aligned}
f(y_1, \boldsymbol{y}_0) =& \sum_{m=1}^{M_1} \alpha_m n_{d(p+1)}((y_1, \boldsymbol{y}_0);\boldsymbol{1}_{p+1}\otimes \mu_m;\Sigma_{m,p+1}) \\
&+ \sum_{m=M_1+1}^M \alpha_m t_{d(p+1)}((y_1, \boldsymbol{y}_0);\boldsymbol{1}_{p+1}\otimes \mu_m;\Sigma_{m,p+1},\nu_m).
\end{aligned}
\end{align}
Integrating $y_{-p+1}$ out, and using the properties of marginal distributions of a multivariate Gaussian and $t$-distributions (see Appendix~\ref{ap:propt}) together with the block Toeplitz form of $\Sigma_{m,p+1}$ shows that the density of $\boldsymbol{y}_1$ is $f(\boldsymbol{y}_1;\boldsymbol{\theta})= \sum_{m=1}^{M_1}\alpha_mn_{dp}(\boldsymbol{y}_1;\boldsymbol{1}_p\otimes\mu_{m},\Sigma_{m,p}) + \sum_{m=M_1+1}^M\alpha_mt_{dp}(\boldsymbol{y}_1;\boldsymbol{1}_p\otimes\mu_{m},\Sigma_{m,p},\nu_m)$. Thus,  $\boldsymbol{y}_0$ and $\boldsymbol{y}_1$ are identically distributed.  As $\lbrace \boldsymbol{y}_t \rbrace_{t=1}^{\infty}$ is a (time-homogeneous) Markov chain, it follows that $\lbrace \boldsymbol{y}_t \rbrace_{t=1}^{\infty}$ has a stationary distribution, say $\pi_{\boldsymbol{y}}(\cdot)$, characterized by the density $f(\cdot;\boldsymbol{\theta})=  \sum_{m=1}^{M_1}\alpha_m n_{dp}(\cdot;\boldsymbol{1}_p\otimes\mu_{m},\Sigma_{m,p}) + \sum_{m=M_1+1}^M\alpha_mt_{dp}(\cdot;\boldsymbol{1}_p\otimes\mu_{m},\Sigma_{m,p},\nu_m)$ \citep[pp. 230-231]{Meyn+Tweedie:2009}.

For ergodicity, let $P_{\boldsymbol{y}}(\boldsymbol{y},\cdot)=\mathbb{P}(\boldsymbol{y}_p\in\cdot|\boldsymbol{y}_0=\boldsymbol{y})$ signify the $p$-step transition probability measure of the process $\boldsymbol{y}_t$. Using the $p$th order Markov property of $y_t$, it is straightforward to check that $P_{\boldsymbol{y}}(\boldsymbol{y},\cdot)$ has the density 
\begin{align}
\begin{aligned}
&f(\boldsymbol{y}_p|\boldsymbol{y}_0;\boldsymbol{\theta})=\prod_{t=1}^pf(y_t|\boldsymbol{y}_{t-1};\boldsymbol{\theta})=\\
&\prod_{t=1}^p\left(\sum_{m=1}^{M_1}\alpha_m n_d(y_1;\mu_{m,t}(\boldsymbol{y}_{t-1}),\Omega_{m}) +  \sum_{m=M_1 + 1}^M\alpha_m t_d(y_1;\mu_{m,t}(\boldsymbol{y}_{t-1}),\Omega_{m,t}(\boldsymbol{y}_{t-1}),\nu_m+dp) \right).
\end{aligned}
\end{align}
Clearly, $f(\boldsymbol{y}_p|\boldsymbol{y}_0;\boldsymbol{\theta})>0$ for all $\boldsymbol{y}_0\in\mathbb{R}^{dp}$ and $\boldsymbol{y}_p\in\mathbb{R}^{dp}$, so it can be concluded that $\boldsymbol{y}_t$ is ergodic in the sense of \cite[Chapter~13]{Meyn+Tweedie:2009} by using arguments identical to those used in the proof of Theorem~1 in \cite{Kalliovirta+Meitz+Saikkonen:2015}.$\blacksquare$

\subsection{Proof of Theorem~\ref{thm:mle}}\label{sec:thmmleproof}
First note that $L_T^{(c)}(\boldsymbol{\theta})$ is continuous and that together with Assumption~\ref{as:mle} it implies existence of a measurable maximizer $\hat{\boldsymbol{\theta}}_T$. To conclude that $\hat{\boldsymbol{\theta}}_T$ is strongly consistent, we need to show that \citep[see, e.g., ][Theorem 2.1 and the discussion on page 2122]{Newey+McFadden:1994}
\begin{enumerate}[label=(\roman*)]
\item the uniform strong law of large numbers holds for the log-likelihood function; that is, \\
$$
\underset{\boldsymbol{\theta}\in\boldsymbol{\Theta}}{\sup}\left| L_T^{(c)}(\boldsymbol{\theta}) - E[L_T^{(c)}(\boldsymbol{\theta})] \right|\rightarrow 0 \quad \text{almost surely as} \quad T\rightarrow\infty,
$$\label{enumProofThm3:USLL}
\item and that the limit of $L_T^{(c)}(\boldsymbol{\theta})$ is uniquely maximized at $\boldsymbol{\theta} = \boldsymbol{\theta}_0$.\label{enumProofThm3:ident}
\end{enumerate}

\textbf{Proof of \ref{enumProofThm3:USLL}}. By Theorem~\ref{thm:statdist}, the process $\boldsymbol{y}_{t-1}=(y_t,...,y_{t-p+1})$, and hence also $y_t$, is stationary and ergodic, and $E[L_T^{(c)}(\boldsymbol{\theta})]=E[l_t(\boldsymbol{\theta})]$. To conclude \ref{enumProofThm3:USLL}, it therefore suffices to show that $E[\sup_{\boldsymbol{\theta}\in\boldsymbol{\Theta}}|l_t(\boldsymbol{\theta})|]<\infty$ \citep[see][]{RangaRao:1962}.  
We will do that by making use of the compactness of the parameter space to derive finite lower and upper bounds for $l_t(\boldsymbol{\theta})$, which is given as
\begin{equation}\label{eq:l_t}
l_t(\boldsymbol{\theta}) = \log\left(\sum_{m=1}^{M_1}  \alpha_{m,t} n_d(y_t;\mu_{m,t},\Omega_{m}) + \sum_{m=M_1+1}^M  \alpha_{m,t}t_d(y_t;\mu_{m,t},\Omega_{m,t},\nu_m + dp) \right).
\end{equation}

First, we derive an upper bound for the normal distribution densities. Determinant of the positive definite conditional covariance matrix $\Omega_{m}$ is a continuous function of the parameters $vech(\Omega_m)$, and hence, compactness of the parameter space implies that the determinant is bounded from below by some constant that is strictly larger than zero and from above by some finite constant. Thus, 
\begin{equation}\label{eq:det_omega_bounds}
0<c_1 \leq \det(\Omega_{m})^{-1/2} \leq c_2 <\infty,
\end{equation}
for some constants $c_1$ and $c_2$. Because $\Omega_m^{-1}$ is positive definite and exponential function is bounded from above by one in the non-positive real axis, we obtain the upper bound
\begin{equation}
n_d(y_t;\mu_{m,t},\Omega_m) = (2\pi)^{-d/2}\text{det}(\Omega_m)^{-1/2}\exp\left\lbrace -\frac{1}{2}(y_t -\mu_m)'\Omega_m^{-1}(y_t - \mu_m) \right\rbrace \leq  (2\pi)^{-d/2}c_2.
\end{equation}

Next, we derive an upper bound for the $t$-distribution densities 
\begin{align}
\begin{aligned}
t_d(y_t;\mu_{m,t},\Omega_{m,t},\nu_m + dp) =& \frac{\Gamma\left(\frac{\nu_m+ (1+ p)d}{2}\right)}{\sqrt{\pi^d(\nu_m+ dp-2)^d}\Gamma\left(\frac{\nu_m+ dp}{2}\right)}\det (\Omega_{m,t})^{-1/2} \\
&\times\left(1 + \frac{(y_t - \mu_{m,t})'\Omega_{m,t}^{-1}(y_t - \mu_{m,t})}{\nu_m + dp - 2} \right)^{-(\nu_m + d(1 + p))/2}.
\end{aligned}
\end{align}
Since $\nu_m > 2$ and the parameter space is compact, $2<c_3\leq\nu_m \leq c_4<\infty$ for some constants $c_3$ and $c_4$. Because the gamma function is continuous on the positive real axis, it then follows that 
\begin{equation}\label{eq:C_bounds}
0<c_5\leq \frac{\Gamma\left(\frac{\nu_m+ (1+ p)d}{2}\right)}{\sqrt{\pi^d(\nu_m+ dp-2)^d}\Gamma\left(\frac{\nu_m+ dp}{2}\right)} \leq c_6
\end{equation}
for some finite constants $c_5$ and $c_6$.  

Using the bounds $2<c_3\leq\nu_m \leq c_4<\infty$ and (\ref{eq:det_omega_bounds}) together with the fact that $\Sigma_{m,p}^{-1}$ is positive definite gives
\begin{align}
\begin{aligned}
\det(\Omega_{m,t})^{-1/2} &= \left(\frac{\nu_m - 2 + (\boldsymbol{y}_{t-1} - \boldsymbol{1}_p\otimes\mu_m)'\Sigma_{m,p}^{-1}(\boldsymbol{y}_{t-1} - \boldsymbol{1}_p\otimes\mu_m)}{\nu_m - 2 + dp}\right)^{-d/2} \det(\Omega_m)^{-1/2}\\
&\leq \left(\frac{c_3 - 2}{c_4 + dp - 2}\right)^{-d/2}c_2 <\infty.
\end{aligned}
\end{align}
For a lower bound, note that $\Sigma_{m,p}^{-1}$ is a continuous function of the parameters and thereby its eigenvalues are as well.  It then follows from the compactness of the parameter space that its largest eigenvalue,  $\lambda_1^{max}$, is bounded from above by some finite constant, say $c_7$. The compactness of the parameter space also implies that there exist finite constant $c_8$ such that $\mu_{im}\leq c_8$ for all $i=1,..,d$ (where $\mu_{im}$ is the $i$th element of $\mu_{m}$).  By using the orthonormal spectral decomposition of $\Sigma_{m,p}^{-1}$, we then obtain
\begin{align}
\begin{aligned}
(\boldsymbol{y}_{t-1} - \boldsymbol{1}_p\otimes\mu_m)'\Sigma_{m,p}^{-1}(\boldsymbol{y}_{t-1} - \boldsymbol{1}_p\otimes\mu_m) &\leq \lambda_1^{max}(\boldsymbol{y}_{t-1} - \boldsymbol{1}_p\otimes\mu_m)'(\boldsymbol{y}_{t-1} - \boldsymbol{1}_p\otimes\mu_m) \\
&\leq c_7(\boldsymbol{y}_{t-1}'\boldsymbol{y}_{t-1} - 2c_8\boldsymbol{y}_{t-1}'\boldsymbol{1}_{dp} + dpc_8^2).
\end{aligned}
\end{align}
Thus,
\begin{equation}\label{eq:det_Omega_mt_lowerbound}
\det(\Omega_{m,t})^{-1/2} \geq \left(\frac{c_4 - 2 + c_7(\boldsymbol{y}_{t-1}'\boldsymbol{y}_{t-1} - 2c_8\boldsymbol{y}_{t-1}'\boldsymbol{1}_{dp} + dpc_8^2)}{c_3 - 2 + dp}\right)^{-d/2}c_1.
\end{equation}

As $-(\nu_m + (1 + p)d)/2<0$ and $\Omega_{m,t}^{-1}$ is positive definite, we have that 
\begin{equation}
 \left(1+\frac{(y_t -\mu_{m,t})'\Omega_{m,t}^{-1}(y_t - \mu_{m,t})}{\nu_m+ dp-2}\right)^{-(\nu_m + (1 + p)d)/2}\leq 1.
\end{equation}
Hence, $t_d(y_t;\mu_{m,t},\Omega_{m,t},\nu_m + dp)\leq \left(\frac{c_3 - 2}{c_4 + dp - 2}\right)^{-d/2}c_2c_6$. It then follows from $\sum_{m=1}^{M}\alpha_{m,t}=1$ that 
\begin{equation}\label{eq:l_t_upperbound}
l_t(\boldsymbol{\theta}) \leq \log\left(\max\left\lbrace (2\pi)^{-d/2}c_2, \left(\frac{c_3 - 2}{c_4 + dp - 2}\right)^{-d/2}c_2c_6\right\rbrace\right) <\infty.
\end{equation}
That is, $l_t(\boldsymbol{\theta})$ is bounded from above by a finite constant.

Next, we proceed by bounding $l_t(\boldsymbol{\theta})$ from below.  Since the eigenvalues of $\Omega_{m}^{-1}$ are continuous functions of the parameters bounded by compactness of the parameter space, the largest eigenvalue, $\lambda_2^{max}$, is bounded from above by some finite constant, say $c_9$. Making use of the orthonormal spectral decomposition of $\Omega_{m}^{-1}$, we then obtain
\begin{align}
\begin{aligned}\label{eq:quadratic_upper1}
(y_t -\mu_{m,t})'\Omega_{m}^{-1}(y_t - \mu_{m,t}) & \leq \lambda_2^{max}(y_t -\boldsymbol{A}_{m,p}\boldsymbol{y}_{t-1})'(y_t - \boldsymbol{A}_{m,p}\boldsymbol{y}_{t-1})\\
& \leq c_9(y_t'y_t - 2y_t'\boldsymbol{A}_{m,p}\boldsymbol{y}_{t-1} + \boldsymbol{y}_{t-1}'\boldsymbol{A}_{m,p}'\boldsymbol{A}_{m,p}\boldsymbol{y}_{t-1}).
\end{aligned}
\end{align}
The compactness of the parameter space implies that 
\begin{equation}
\boldsymbol{y}_{t-1}'\boldsymbol{A}_{m,p}'\boldsymbol{A}_{m,p}\boldsymbol{y}_{t-1} \leq c_{10} \sum_{i=1}^{dp}\sum_{j=1}^{dp}|\boldsymbol{y}_{j,t-1}\boldsymbol{y}_{i,t-1}| 
\end{equation}
for some finite constant $c_{10}$, where $\boldsymbol{y}_{i,t-1}$ is the $i$th element of $\boldsymbol{y}_{t-1}$.
Denoting by $a_{m,i}(k,j)$ the $kj$th element of the autoregression matrix $A_{m,i}$ and $y_{kt}$ the $k$th element of $y_t$, we have
\begin{equation}
y_t'\boldsymbol{A}_{m,p}\boldsymbol{y}_{t-1} = \sum_{k=1}^d\sum_{i=1}^p\sum_{j=1}^da_{m,i}(k,j)y_{kt}y_{jt-i}\leq  \sum_{k=1}^d\sum_{i=1}^p\sum_{j=1}^dc_{11}|y_{kt}y_{jt-i}|,
\end{equation}
where $c_{11}$ is a finite constant that bounds the absolute values of the autoregression coefficients from above (which exists due to compactness of the parameter space). Combining the above two bounds with (\ref{eq:quadratic_upper1}) gives the upper bound
\begin{equation}\label{eq:qformbound}
(y_t -\mu_{m,t})'\Omega_{m}^{-1}(y_t - \mu_{m,t}) \leq c_{12}\left(y_t'y_t + \sum_{i=1}^{dp}\sum_{j=1}^{dp}|\boldsymbol{y}_{j,t-1}\boldsymbol{y}_{i,t-1}|  + \sum_{k=1}^d\sum_{i=1}^p\sum_{j=1}^d|y_{kt}y_{jt-i}|\right).
\end{equation}
where $c_{12}$ is a finite constant. 

Using the fact that $\Sigma_{m,p}^{-1}$ is positive definite together with the bounds $2<c_3\leq\nu_m \leq c_4<\infty$ shows that 
\begin{align}
\begin{aligned}
\Omega_{m,t}^{-1} &= \frac{\nu_m - 2 + dp}{\nu_m - 2 + (\boldsymbol{y}_{t-1} - \boldsymbol{1}_p\otimes\mu_m)'\Sigma_{m,p}^{-1}(\boldsymbol{y}_{t-1} - \boldsymbol{1}_p\otimes\mu_m)}\Omega_m^{-1}\leq \frac{c_4 - 2 + dp}{c_3 - 2}\Omega_m^{-1}.
\end{aligned}
\end{align}
Using the above inequality together with $2<c_3\leq\nu_m$ and (\ref{eq:qformbound}) then gives
\begin{equation}\label{eq:t_qformthing_upperbound}
\frac{(y_t -\mu_{m,t})'\Omega_{m,t}^{-1}(y_t - \mu_{m,t})}{v_m + pd - 2} \leq c_{13}\left(y_t'y_t + \sum_{i=1}^{dp}\sum_{j=1}^{dp}|\boldsymbol{y}_{j,t-1}\boldsymbol{y}_{i,t-1}| + \sum_{k=1}^d\sum_{i=1}^p\sum_{j=1}^d|y_{kt}y_{jt-i}|\right),
\end{equation}
where $c_{13} = ((c_3 - 2)(c_3 + pd - 2))^{-1}(c_4-2+dp)c_{12}$ is a finite constant.

From $\sum_{m=1}^{M}\alpha_{m,t}=1$, (\ref{eq:det_omega_bounds}), (\ref{eq:C_bounds}), (\ref{eq:det_Omega_mt_lowerbound}), (\ref{eq:qformbound}), (\ref{eq:t_qformthing_upperbound}), and $\nu_m\leq c_4$, we then obtain a lower bound for $l_t(\boldsymbol{\theta})$ as
\begin{align}\label{eq:l_t_lowerbound}
\begin{aligned}
l_t(\boldsymbol{\theta})  \geq \min&\left\lbrace -\frac{d}{2}\log(2\pi) + \log(c_1)\right. \\
&- \frac{1}{2}c_{12}\left(y_t'y_t + \sum_{i=1}^{dp}\sum_{j=1}^{dp}|\boldsymbol{y}_{j,t-1}\boldsymbol{y}_{i,t-1}| + \sum_{k=1}^d\sum_{i=1}^p\sum_{j=1}^d|y_{kt}y_{jt-i}|\right),\\
& c_{15} - \frac{d}{2}\log (c_4 - 2 + c_7(\boldsymbol{y}_{t-1}'\boldsymbol{y}_{t-1} - 2c_8\boldsymbol{y}_{t-1}'\boldsymbol{1}_{dp} + dpc_8^2))\\ 
&\left.- c_{14}\log\left(1 +  c_{13}\left(y_t'y_t +\sum_{i=1}^{dp}\sum_{j=1}^{dp}|\boldsymbol{y}_{j,t-1}\boldsymbol{y}_{i,t-1}| + \sum_{k=1}^d\sum_{i=1}^p\sum_{j=1}^d|y_{kt}y_{jt-i}|\right)\right)\right\rbrace,
\end{aligned}
\end{align}
where $c_{14}= (c_4 + (1 + p)d)/2$ and $c_{15}=\log(c_5) + \log(c_1) + \frac{d}{2}(c_3 - 2 + dp)$.
Since $y_t$ is stationary with finite second moments, it holds that
\begin{align}\label{eq:finitemoment}
\begin{aligned}
&E\left[y_t'y_t + \sum_{i=1}^{dp}\sum_{j=1}^{dp}|\boldsymbol{y}_{j,t-1}\boldsymbol{y}_{i,t-1}| + \sum_{k=1}^d\sum_{i=1}^p\sum_{j=1}^d|y_{kt}y_{jt-i}|\right]<\infty  \ \ \text{and} \\
&E[\boldsymbol{y}_{t-1}'\boldsymbol{y}_{t-1} - 2c_8\boldsymbol{y}_{t-1}'\boldsymbol{1}_{dp}]< \infty,
\end{aligned}
\end{align}
and thereby we  obtain from Jensen's inequality that also
\begin{align}\label{eq:finitelogmoment}
\begin{aligned}
&E\left[\log\left(1 +  c_{13}\left(y_t'y_t + \sum_{i=1}^{dp}\sum_{j=1}^{dp}|\boldsymbol{y}_{j,t-1}\boldsymbol{y}_{i,t-1}| + \sum_{k=1}^d\sum_{i=1}^p\sum_{j=1}^d|y_{kt}y_{jt-i}|\right)\right)\right]<\infty \ \ \text{and}\\
& E[\log (c_4 - 2 + c_7(\boldsymbol{y}_{t-1}'\boldsymbol{y}_{t-1} - 2c_8\boldsymbol{y}_{t-1}'\boldsymbol{1}_{dp} + dpc_8^2))]<\infty .
\end{aligned}
\end{align}
The upper bound (\ref{eq:l_t_upperbound}) together with (\ref{eq:l_t_lowerbound}), (\ref{eq:finitemoment}), and (\ref{eq:finitelogmoment}) shows that $E[\sup_{\boldsymbol{\theta}\in\boldsymbol{\Theta}}|l_t(\boldsymbol{\theta})|]<\infty$.$\blacksquare$

\textbf{Proof of \ref{enumProofThm3:ident}}. To prove that $E[l_t(\boldsymbol{\theta})]$ is uniquely maximized at $\boldsymbol{\theta}=\boldsymbol{\theta}_0$, it needs to be shown that $E[l_t(\boldsymbol{\theta})]\leq E[l_t(\boldsymbol{\theta}_0)]$, and that $E[l_t(\boldsymbol{\theta})]= E[l_t(\boldsymbol{\theta}_0)]$ implies 
\begin{align}
\begin{aligned}
& \boldsymbol{\vartheta}_m=\boldsymbol{\vartheta}_{\tau(m),0} \ \ \text{and} \  \ \alpha_{m} = \alpha_{\tau(m),0}  \ \ \text{when}\ \ m=1,....,M_1, \ \ \text{and} \\
& (\boldsymbol{\vartheta}_m,\nu_m) = (\boldsymbol{\vartheta}_{\tau(m),0},  \nu_{\tau(m),0}) \ \ \text{and} \  \ \alpha_{m} = \alpha_{\tau(m),0}  \ \ \text{when}\ \ m=M_1+1,....,M,
\end{aligned}
\end{align}
for some permutations $\lbrace \tau_1(1),...,\tau_1(M_1) \rbrace$ and $\lbrace \tau_2(M_1+1),...,\tau_2(M) \rbrace$. For notational clarity, we write $\mu_{m,t}=\mu(\boldsymbol{y};\boldsymbol{\vartheta}_m)$,  $\Omega_{m}=\Omega(\boldsymbol{\vartheta}_m)$,  $\Omega_{m,t}=\Omega(\boldsymbol{y};\boldsymbol{\vartheta}_m,\nu_m)$, and $\alpha_{m,t}=\alpha_m(\boldsymbol{y};\boldsymbol{\theta})$, making clear their dependence on the parameter value. 

The density of $(y_t,\boldsymbol{y}_{t-1})$ can be written as 
\begin{align}
\begin{aligned}
f((y_t,\boldsymbol{y}_{t-1});\boldsymbol{\theta}_0) =& \sum_{n=1}^{M}\alpha_{n,0}d_{n,dp}(\boldsymbol{y}_{t-1};\boldsymbol{1}_p\otimes\mu_{n,0},\Sigma_{n,p,0},\nu_{n,0}) \times \\
&\left(\sum_{m=1}^{M_1}\alpha_m(\boldsymbol{y};\boldsymbol{\theta}_0)n_d(y_t;\mu(\boldsymbol{y};\boldsymbol{\vartheta}_{m,0}),\Omega(\boldsymbol{\vartheta}_{m,0})) \right. + \\ 
&\left.\sum_{m=M_1+1}^M\alpha_m(\boldsymbol{y};\boldsymbol{\theta}_0)t_d(y_t;\mu(\boldsymbol{y};\boldsymbol{\vartheta}_{m,0}),\Omega(\boldsymbol{y};\boldsymbol{\vartheta}_{m,0},\nu_{m,0}),\nu_{m,0}+dp)\right),
\end{aligned}
\end{align}
where $d_{n,dp}(\cdot;\boldsymbol{1}_p\otimes\mu_{n,0},\Sigma_{n,p,0},\nu_{n,0})$ is defined in (\ref{eq:d_mdp}).  By using this together with reasoning based on Kullback-Leibler divergence, arguments analogous to those in \citet[pp. 494-495]{Kalliovirta+Meitz+Saikkonen:2016} can be used to conclude that $E[l_t(\boldsymbol{\theta})] - E[l_t(\boldsymbol{\theta}_0)] \leq 0$, with equality if and only if for almost all $(y,\boldsymbol{y})\in\mathbb{R}^{d(p+1)}$,
\begin{align}\label{eq:identequality1}
\begin{aligned}
&\sum_{m=1}^{M_1}\alpha_m(\boldsymbol{y};\boldsymbol{\theta})n_d(y_t;\mu(\boldsymbol{y};\boldsymbol{\vartheta}_{m}),\Omega(\boldsymbol{\vartheta}_{m}))+\\
&\sum_{m=M_1+1}^M\alpha_m(\boldsymbol{y};\boldsymbol{\theta})t_d(y_t;\mu(\boldsymbol{y};\boldsymbol{\vartheta}_{m}),\Omega(\boldsymbol{y};\boldsymbol{\vartheta}_{m},\nu_{m}),\nu_{m}+dp)\\
& =\sum_{m=1}^{M_1}\alpha_m(\boldsymbol{y};\boldsymbol{\theta}_0)n_d(y_t;\mu(\boldsymbol{y};\boldsymbol{\vartheta}_{m,0}),\Omega(\boldsymbol{\vartheta}_{m,0}))+\\
&\sum_{m=M_1+1}^M\alpha_m(\boldsymbol{y};\boldsymbol{\theta}_0)t_d(y_t;\mu(\boldsymbol{y};\boldsymbol{\vartheta}_{m,0}),\Omega(\boldsymbol{y};\boldsymbol{\vartheta}_{m,0},\nu_{m,0}),\nu_{m,0}+dp).
\end{aligned}
\end{align}
For each fixed $\boldsymbol{y}$ at a time, the mixing weights, conditional means, and conditional covariances in (\ref{eq:identequality1}) are constants, so we may apply the result on identification of finite mixtures of multivariate Gaussian and $t$-distributions in \citet[Example~1]{Holzmann+Munk+Gneiting:2006} (their parametrization of the $t$-distribution slightly differs from ours, but identification with their parametrization implies identification with our parametrization). For each fixed $\boldsymbol{y}$, there thus exists a permutations $\lbrace \tau_1(1),...,\tau_1(M_1) \rbrace$ and $\lbrace \tau_2(M_1+1),...,\tau_2(M) \rbrace$ (that may depend on $\boldsymbol{y}$) of the index sets $\lbrace 1,...,M_1 \rbrace$ and $\lbrace M_1+1,...,M \rbrace$ such that
\begin{align}\label{eq:perm_alphaetc_n}
\begin{aligned}
&\alpha_m(\boldsymbol{y};\boldsymbol{\theta})=\alpha_{\tau_1(m)}(\boldsymbol{y};\boldsymbol{\theta}_0), \  \mu(\boldsymbol{y};\boldsymbol{\vartheta}_{m}) = \mu(\boldsymbol{y};\boldsymbol{\vartheta}_{\tau_1(m),0}), \ \text{and} \ \Omega(\boldsymbol{\vartheta}_{m}) = \Omega(\boldsymbol{\vartheta}_{\tau_1(m),0}),
\end{aligned}
\end{align}
for $m=1,...,M_1$ and almost all $y\in\mathbb{R}^d$, and
\begin{align}\label{eq:perm_alphaetc_t}
\begin{aligned}
&\alpha_m(\boldsymbol{y};\boldsymbol{\theta})=\alpha_{\tau_2(m)}(\boldsymbol{y};\boldsymbol{\theta}_0), \  \mu(\boldsymbol{y};\boldsymbol{\vartheta}_{m}) = \mu(\boldsymbol{y};\boldsymbol{\vartheta}_{\tau_2(m),0}),  \Omega(\boldsymbol{y};\boldsymbol{\vartheta}_{m}) = \Omega(\boldsymbol{y};\boldsymbol{\vartheta}_{\tau_2(m),0}),\\
& \text{and} \ \ \nu_m = \nu_{\tau_2(m),0}
\end{aligned}
\end{align}
for $m=M_1+1,...,M$ and almost all $y\in\mathbb{R}^d$. Note that from (\ref{eq:perm_alphaetc_n}) we readily obtain $vech(\Omega_m)=vech(\Omega_{\tau_1(m),0})$.

Arguments analogous to those in \citet[p. 495]{Kalliovirta+Meitz+Saikkonen:2016} can then be used to conclude from (\ref{eq:perm_alphaetc_n}) and (\ref{eq:perm_alphaetc_t}) that $\alpha_m=\alpha_{\tau_1(m),0}, \phi_{m,0}=\phi_{\tau_1(m),0,0}$‚ and $\boldsymbol{A}_{m,p}=\boldsymbol{A}_{\tau_1(m),p,0}$ for $m=1,...,M_1$, and $\alpha_m=\alpha_{\tau_2(m),0}, \phi_{m,0}=\phi_{\tau_2(m),0,0}$‚ and $\boldsymbol{A}_{m,p}=\boldsymbol{A}_{\tau_2(m),p,0}$ for $m=M_1+1,...,M$. Given these identities and $\nu_m=\nu_{\tau_2(m),0}$, we obtain from $\Omega(\boldsymbol{y};\boldsymbol{\vartheta}_{m}) = \Omega(\boldsymbol{y};\boldsymbol{\vartheta}_{\tau_2(m),0})$ in (\ref{eq:perm_alphaetc_t}) that
\begin{align}\label{eq:omegas1} 
\begin{aligned}
& (\boldsymbol{y} - \boldsymbol{1}_p\otimes\mu_{\tau_2(m),0})'\Sigma_p(\boldsymbol{\vartheta}_{m})^{-1}(\boldsymbol{y} - \boldsymbol{1}_p\otimes\mu_{\tau_2(m),0})\Omega_m -\\
& (\boldsymbol{y} - \boldsymbol{1}_p\otimes\mu_{\tau_2(m),0})'\Sigma_p(\boldsymbol{\vartheta}_{\tau_2(m),0})^{-1}(\boldsymbol{y} - \boldsymbol{1}_p\otimes\mu_{\tau_2(m),0})\Omega_{\tau_2(m),0}
 =  (\nu_{\tau_2(m),0} - 2)(\Omega_{\tau_2(m),0} - \Omega_m).
 \end{aligned}
\end{align}
The condition $\Omega(\boldsymbol{y};\boldsymbol{\vartheta}_{m}) = \Omega(\boldsymbol{y};\boldsymbol{\vartheta}_{\tau_2(m),0})$ implies that $\Omega_m$ is proportional to $\Omega_{\tau_2(m),0}$,  say $\Omega_m = c(\boldsymbol{\vartheta}_{m,\tau_2(m)}^+) \Omega_{\tau_2(m),0}$, where the strictly positive scalar $c(\boldsymbol{\vartheta}_{m,\tau_2(m)}^+)$ may depend on the parameter $\boldsymbol{\vartheta}_{m,\tau_2(m)}^+ \equiv (\boldsymbol{\vartheta}_m,\boldsymbol{\vartheta}_{\tau_2(m),0},\nu_{\tau_2(m),0})$. It is then easy to see from the vectorized structure of $\Sigma_p(\cdot)$, given in (\ref{eq:gausquantities}), that $\Sigma_p(\boldsymbol{\vartheta}_{m})^{-1}=c(\boldsymbol{\vartheta}_{m,\tau_2(m)}^+)^{-1}\Sigma_p(\boldsymbol{\vartheta}_{\tau_2(m),0})^{-1}$. By using this together with  the identity $\Omega_m = c(\boldsymbol{\vartheta}_{m,\tau_2(m)}^+) \Omega_{\tau_2(m),0}$, the left hand side of (\ref{eq:omegas1}) reduces to
\begin{align}
\begin{aligned}
& (\boldsymbol{y} - \boldsymbol{1}_p\otimes\mu_{\tau_2(m),0})'(c(\boldsymbol{\vartheta}_{m,\tau_2(m)}^+)\Sigma_p(\boldsymbol{\vartheta}_m)^{-1} - \Sigma_p(\boldsymbol{\vartheta}_{\tau_2(m),0})^{-1})(\boldsymbol{y} - \boldsymbol{1}_p\otimes\mu_{\tau_2(m),0})\Omega_{\tau_2(m),0}  \\
& =(\boldsymbol{y} - \boldsymbol{1}_p\otimes\mu_{\tau_2(m),0})'\left(\frac{c(\boldsymbol{\vartheta}_{m,\tau_2(m)}^+)}{c(\boldsymbol{\vartheta}_{m,\tau_2(m)}^+)}\Sigma_p(\boldsymbol{\vartheta}_{\tau_2(m),0})^{-1} - \Sigma_p(\boldsymbol{\vartheta}_{\tau_2(m),0})^{-1}\right)(\boldsymbol{y} - \boldsymbol{1}_p\otimes\mu_{\tau_2(m),0})\\
&\times\Omega_{\tau_2(m),0}=0.
\end{aligned}
\end{align}
Thereby (\ref{eq:omegas1}) reduces to $(\nu_{\tau_2(m),0} - 2)(\Omega_{\tau_2(m),0} - \Omega_m)=0$,  which implies $\Omega_m = \Omega_{\tau_2(m),0}$, as $\nu_{\tau_2(m),0} > 2$. Since the condition (\ref{eq:identcond}) sets a unique ordering for the mixture components, it follows that $\boldsymbol{\theta}=\boldsymbol{\theta}_0$‚ completing the proof of consistency. 

Given consistency and assumptions of the theorem, asymptotic normality of the ML estimator can be concluded using the standard arguments. The required steps can be found, for example, in \citet[proof of Theorem~3]{Kalliovirta+Meitz+Saikkonen:2016}. We omit the details for brevity.$\blacksquare$

\section{Monte Carlo algorithm}\label{sec:montecarlo}
We present a Monte Carlo algorithm that produces point estimates, and with a random initial value $\boldsymbol{y}_{t-1}=(y_{t-1},...,y_{t-p})$, confidence intervals for the generalized impulse response function defined in (\ref{eq:girf}). Our algorithm is adapted from \citet[pp. 135-136]{Koop+Pesaran+Potter:1996} and \citet[pp. 601-602]{Kilian+Lutkepohl:2017}. We assume that the initial value $\boldsymbol{y}_{t-1}$ follows a known distribution $G$, which may be such that it produces a single outcome with probability one (corresponding to a fixed $\boldsymbol{y}_{t-1}$), or it can be the stationary distribution of the process or of a specific regime, for example. In the following, $y_{t+h}^{(j)}(\delta_i,\boldsymbol{y}_{t-1})$ denotes a realization of the process at the time $t+h$ conditional on the structural shock of sign and size $\delta_i$ in the $i$th element of $e_t$ hitting the system at time $t$ and on the $p$ observations $\boldsymbol{y}_{t-1}=(y_{t-1},...,y_{t-p})$ preceding the time $t$, whereas $y_{t+h}^{(j)}(\boldsymbol{y}_{t-1})$ denotes an alternative realization conditional on the initial value $\boldsymbol{y}_{t-1}$ only. The superscript $(j)$ signifies that the realization is related to the $j$th Monte Carlo repetition.

The algorithm proceeds with the following steps.
\begin{enumerate}\addtocounter{enumi}{-1}
\item Decide the horizon $H$, the numbers of repetitions $R_1$ and $R_2$, and the sign and size $\delta_i$ for the $i$th structural shock (that is of interest).

\item Draw an initial value $\boldsymbol{y}_{t-1}$ from $G$.\label{step1}

\item Draw $H+1$ independent realizations of a shock $\varepsilon_t$ from $N(0,I_d)$. For Student's $t$ regimes, the Gaussian shocks are used obtain shocks from the appropriate Student's $t$ distributions. For each of the $M_2$ Student's $t$ regimes, draw also $H+1$ independent realizations of shocks from the $\chi^2_{\nu_m + dp}$-distribution. Moreover, draw an initial regime $m\in \lbrace 1,...,M \rbrace$ according to the probabilities given by the mixing weights $\alpha_{1,t},...,\alpha_{M,t}$, and compute the reduced form shock $u_t=W\Lambda_m^{1/2}\varepsilon_t$, where $\Lambda_1=I_d$. Then, compute the structural shock $e_t = B_t^{-1}u_t$ and impose the size $\delta_i$ on its $i$th element to obtain $e_t^*$. Finally, calculate the modified reduced form shock $u_t^*=B_te_t^*$.\label{step2}

\item Use the modified reduced form shock $u_t^*$ as well as the rest $H$ standard normal shocks $\varepsilon_t$ and draws from $\chi^2_{\nu_m + dp}$-distributions obtained from Step~\ref{step2} to compute realizations $y_{t+h}^{(j)}(\delta_i,\boldsymbol{y}_{t-1})$ for $h=0,1,...,H$, iterating forward so that in each iteration the regime $m$ that generates the observation is first drawn according to the probabilities given by the mixing weights. At $h=0$, the initial regime and the modified reduced form shock $u_t^*$ calculated from the structural shock in Step~\ref{step2} is used. From $h=1$ onwards, the $h+1$th standard normal shock $\varepsilon_t$, and in the case of a Student's $t$ regime also the $h+1$th draw from the $\chi^2_{\nu_m + dp}$-distribution, is used to calculate the reduced form shock $u_{t+h}=W\Lambda_m^{1/2}e_{t+h}$, where $\Lambda_1=I_d$ and $m$ is the selected regime.

\item Use the reduced form shock $u_t$ and the rest $H$ the standard normal shocks $\varepsilon_t$ as well as draws from the $\chi^2_{\nu_m + dp}$-distributions obtained in Step~\ref{step2} to compute realizations $y_{t+h}^{(j)}(\boldsymbol{y}_{t-1})$ for $h=0,1,...,H$ so that the reduced form shock $u_t$ (calculated in Step~\ref{step2}) is used to compute the time $h=0$ realization. Otherwise proceed similarly to the previous step.

\item Calculate $y_{t+h}^{(j)}(\delta_i,\boldsymbol{y}_{t-1}) - y_{t+h}^{(j)}(\boldsymbol{y}_{t-1})$.\label{step5}

\item Repeat Steps~\ref{step2}-\ref{step5} $R_1$ times and calculate the sample mean of $y_{t+h}^{(j)}(\delta_i,\boldsymbol{y}_{t-1}) - y_{t+n}^{(j)}(\boldsymbol{y}_{t-1})$ for $h=0,1,...,H$ to obtain an estimate of the GIRF$(h,\delta_i,\boldsymbol{y}_{t-1})$.\label{step6}

\item Repeat Steps~\ref{step1}-\ref{step6} $R_2$ times to obtain estimates of GIRF$(h,\delta_i,\boldsymbol{y}_{t-1})$ with different starting values $\boldsymbol{y}_{t-1}$ generated from the distribution $G$. Then, take the sample mean and sample quantiles over the estimates to obtain point estimate and confidence intervals for the GIRF with random initial value.\label{step7}
\end{enumerate}
Notice that if a fixed initial value $\boldsymbol{y}_{t-1}$ is used, Step~\ref{step7} is redundant.

\section{Details on the Empirical Application}\label{sec:empdetails}

\subsection{Model selection and adequacy}\label{sec:adequacy}
We estimate the models based on the exact log-likelihood function. The estimation and other numerical analysis is carried out with the accompanying CRAN distributed R package gmvarkit \cite{gmvarkit} to which we have implemented the introduced methods. The R package gmvarkit also contains the data studied in the empirical application to facilitate the reproduction of our results. For evaluating the adequacy of the estimated models, we employ quantile residuals diagnostics in the framework proposed by \cite{Kalliovirta+Saikkonen:2010} \citep[see also the related by][discussing quantile residuals diagnostics in the univariate setting]{Kalliovirta:2012}. For a correctly specified G-StMVAR model, the multivariate quantile residuals are asymptotically standard normally distributed and can thereby be used for graphical diagnostics similarly to the conventional Pearson residuals \citep[Lemma 3]{Kalliovirta+Saikkonen:2010}.\footnote{\cite{Kalliovirta+Saikkonen:2010} also propose formal diagnostic tests for testing normality, autocorrelation, and conditional heteroskedasticity of the quantile residuals. The tests take into account the uncertainty about the true parameter value and can be calculated based on the observed data or by employing a simulation procedure for better size properties. However, we find these tests quite forgiving without the simulation procedure, while with the simulation procedure using a sample of length $10000$, all the tests reject the adequacy our StMVAR model at all the conventional levels of significance. To obtain a better perception of the model's adequacy, we therefore rather employ graphical diagnostics.} For brevity, we show the diagnostic figures for the selected model only.

We start by estimating one-regime StMVAR models with autoregressive orders $p=1,...,12$ and found that BIC, HQIC, and AIC are all minimized by the order $p=1$. Graphical quantile residual diagnostics revealed that the StMVAR($1,1$) model is somewhat inadequate particularly in capturing conditional heteroskedasticity of the series and movements of the interest rate variable, whose quantile residuals' time series displays a shift in volatility and marginal distribution substantial excess kurtosis. Increasing the autoregressive order to $2$ or $3$ does not improve the adequacy much. Moreover, increasing $p$ from $2$ to $3$ decreased the log-likelihood, suggesting that the order $p=3$ might not be suitable for a StMVAR model. The log-likelihoods and values of the information criteria are presented in Table~\ref{tab:ic} for the discussed models.

As the one-regime models are found inadequate, we estimate a two-regime StMVAR model with $p=1$, i.e., a G-StMVAR($p=1,M_1=0,M_2=2$) model. Compared to the one-regime models, particularly the time series plot and marginal distribution of the interest variable's quantile residuals is substantially more reasonable. However, we find that this model has several moderately sized correlation coefficients (CC) at small lags in the autocorrelation function (ACF) of its quantile residuals and squared quantile residuals. To improve the adequacy of the model, we increase the autoregressive order to $p=2$, which decreased many of the moderately sized CCs but increased AIC (see Table~\ref{tab:ic}). The AIC is, nevertheless, smaller than for any of the one-regime models (while the one-regime StMVAR $p=1,2$ models have smaller BIC and the $p=1$ model also smaller HQIC). As is discussed next, we find the overall adequacy of this model reasonable.\footnote{It is possible that superior fitness of the two-regime models is due to the accommodation of regime-switching error covariance matrices or kurtosis and cannot be attributed to the time-varying AR matrices or intercepts. To test whether this is the case, we estimate two restricted StMVAR($2,2$) models. In the first one, we restrict the AR matrices to identical in both regimes, whereas in the second one, we restrict AR matrices and intercepts to be identical in both regimes. Because these models are nested to the unrestricted StMVAR model, the restrictions can be tested with a likelihood ratio test (assuming the validity of the unverified assumptions made in Theorem~\ref{thm:mle}). The former type restrictions obtained the $p$-value $0.022$ and the latter type restrictions the $p$-value $2\cdot 10^{-4}$. We also repeat the exercise for the StMVAR($1,2$) model, which minimized AIC, and obtained the $p$-values $0.003$ and $2\cdot 10^{-7}$ for the restrictions, respectively. As the restrictions are rejected at the $5\%$ level of significance or less, it seems plausible that AR matrices and intercepts vary in time.}

\begin{table}
\centering
\begin{tabular}{c c c c c c c c c c}
Model                               & Log-lik  & BIC     & HQIC    & AIC     \\ 
\hline\\[-1.5ex]
StMVAR($1,1$)                       & $-2.486$ & $5.604$ & $5.360$ & $5.197$ \\
StMVAR($2,1$)                       & $-2.447$ & $5.851$ & $5.482$ & $5.235$ \\
StMVAR($3,1$)                       & $-2.471$ & $6.224$ & $5.729$ & $5.398$ \\
StMVAR($1,2$)                       & $-2.211$ & $5.705$ & $5.210$ & $4.878$ \\
StMVAR($2,2$)                       & $-2.138$ & $6.210$ & $5.464$ & $4.964$ \\
\hline
\end{tabular}
\caption{The log-likelihoods and values of the information criteria divided by the number of observations for the discussed StMVAR($p,M$) models.}
\label{tab:ic}
\end{table}

\begin{figure}[H]
    \centerline{\includegraphics[width=\textwidth - 3cm]{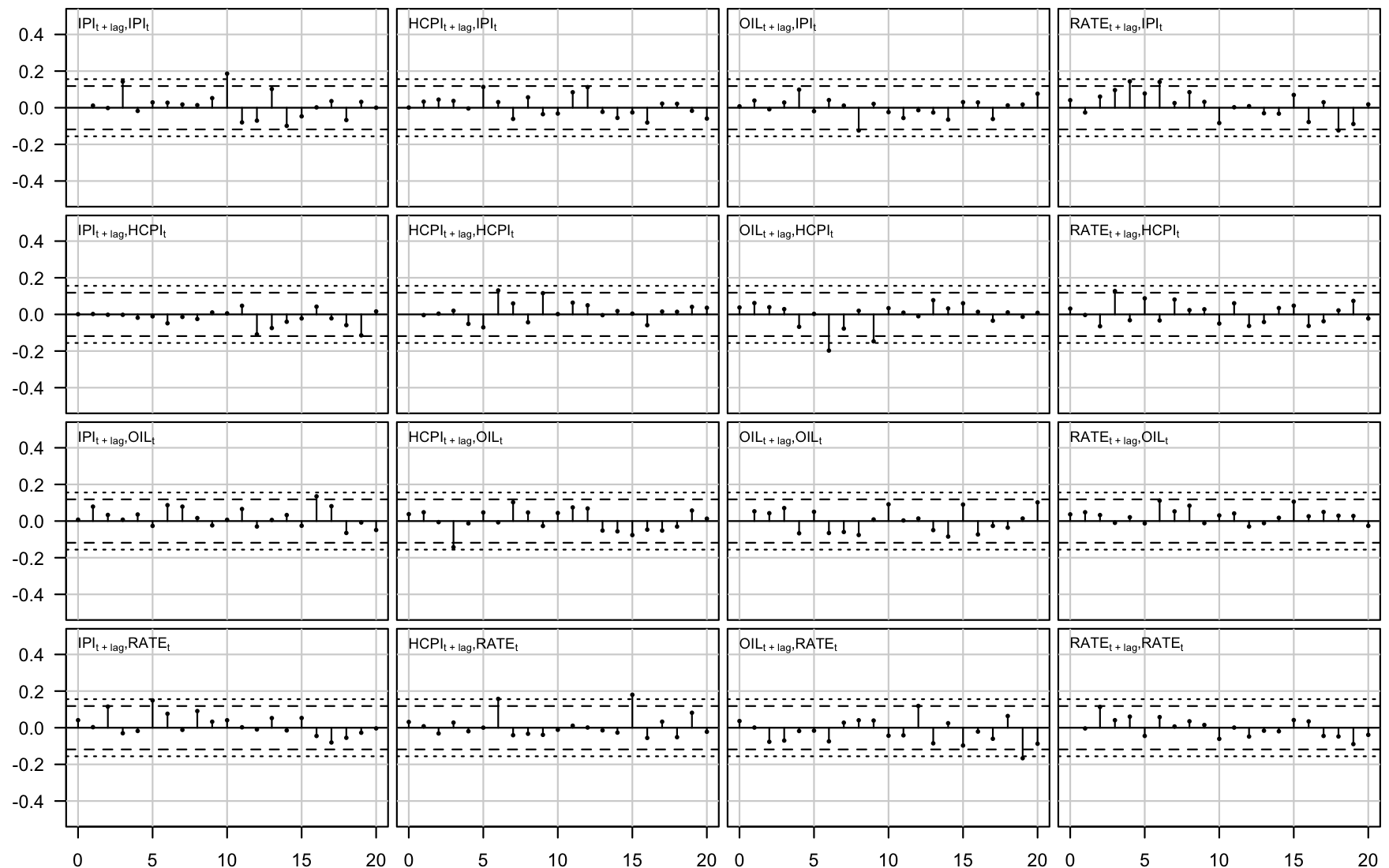}}
    \caption{Auto- and crosscorrelation functions of the quantile residuals of the fitted StMVAR($2,2$) model for the lags $0,1,...,20$. The lag zero autocorrelation coefficients are omitted, as they are one by convention. The dashed lines are the $95\%$ bounds $\pm 1.96/\sqrt{T}$ for autocorrelations of IID observations, whereas the dotted lines are the corresponding $99\%$ bounds $\pm 2.58/\sqrt{T}$.}
\label{fig:acplot}
\end{figure}

Figure~\ref{fig:acplot} presents the ACF and crosscorrelation function (CCF) of the quantile residuals of the StMVAR($2,2$) model for the first $20$ lags. As the figure shows, there is not much autocorrelation in the residuals, but CCs of almost $0.2$ in absolute value stick out in the ACF of IPI's quantile residuals at the lag $10$, in the CCF of OIL and HCPI at the lag $6$, and in the CCF of HCPI and RATE at the lag $15$. There are also a moderately sized CCs at small lags in the ACF of the IPI's quantile residuals at the lag $3$ and in the RATE's quantile residuals at the lag $2$. These CCs are not, however, very large, and as $316$ CCs are presented, some of them are expected to be moderate for a correctly specified model. Therefore, our model appears to capture the autocorrelation structure of the series reasonably well, although some of the CCs are slightly larger than what one would expect for an IID process.

\begin{figure}[H]
    \centerline{\includegraphics[width=\textwidth - 3cm]{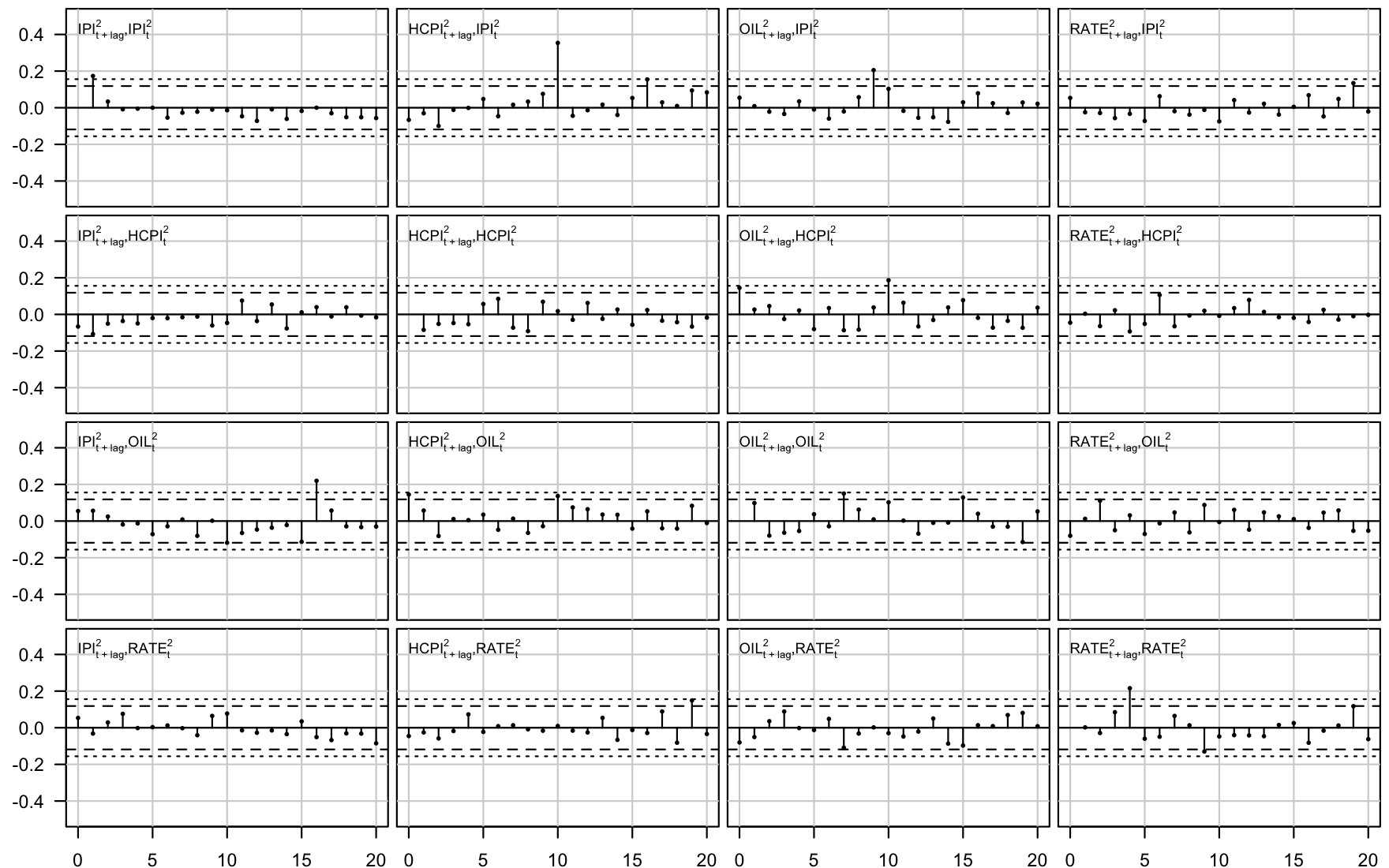}}
    \caption{Auto- and crosscorrelation functions of the squared quantile residuals of the fitted StMVAR($2,2$) model for the lags $0,1,...,20$. The lag zero autocorrelation coefficients are omitted, as they are one by convention. The  dashed lines are the $95\%$ bounds $\pm 1.96/\sqrt{T}$ for autocorrelations of IID observations, whereas the dotted lines are the corresponding $99\%$ bounds $\pm 2.58/\sqrt{T}$.}
\label{fig:chplot}
\end{figure}

The ACF and CCF of the squared quantile residuals are presented in Figure~\ref{fig:chplot} for the first $20$ lags. The figure shows that there is a moderately large CC at the first lag in the ACF of the IPI's squared quantile residuals and a slightly larger one (roughly $0.2$) at the fourth lag in the ACF of RATE's squared quantile residuals. There is a particularly large CC at the lag $10$ in the CCF of HCPI's and IPI's squared quantile residuals, and a somewhat large CC at the lag $16$ in the CCF of IPI's and RATE's squared quantile residuals, at the lag $9$ in the CCF of OIL's and IPI's squared quantile residuals, and at the lag $10$ in the OIL's and HCPI's squared quantile residuals. Nonetheless, the model seems to capture the conditional heteroskedasticity of the series moderately well, as the inadequacies do not seem very severe, with the exception of the single large CC at the lag $10$ in the CCF between squared quantile residuals of HCPI and IPI.

\begin{figure}[H]
    \centerline{\includegraphics[width=\textwidth - 3cm]{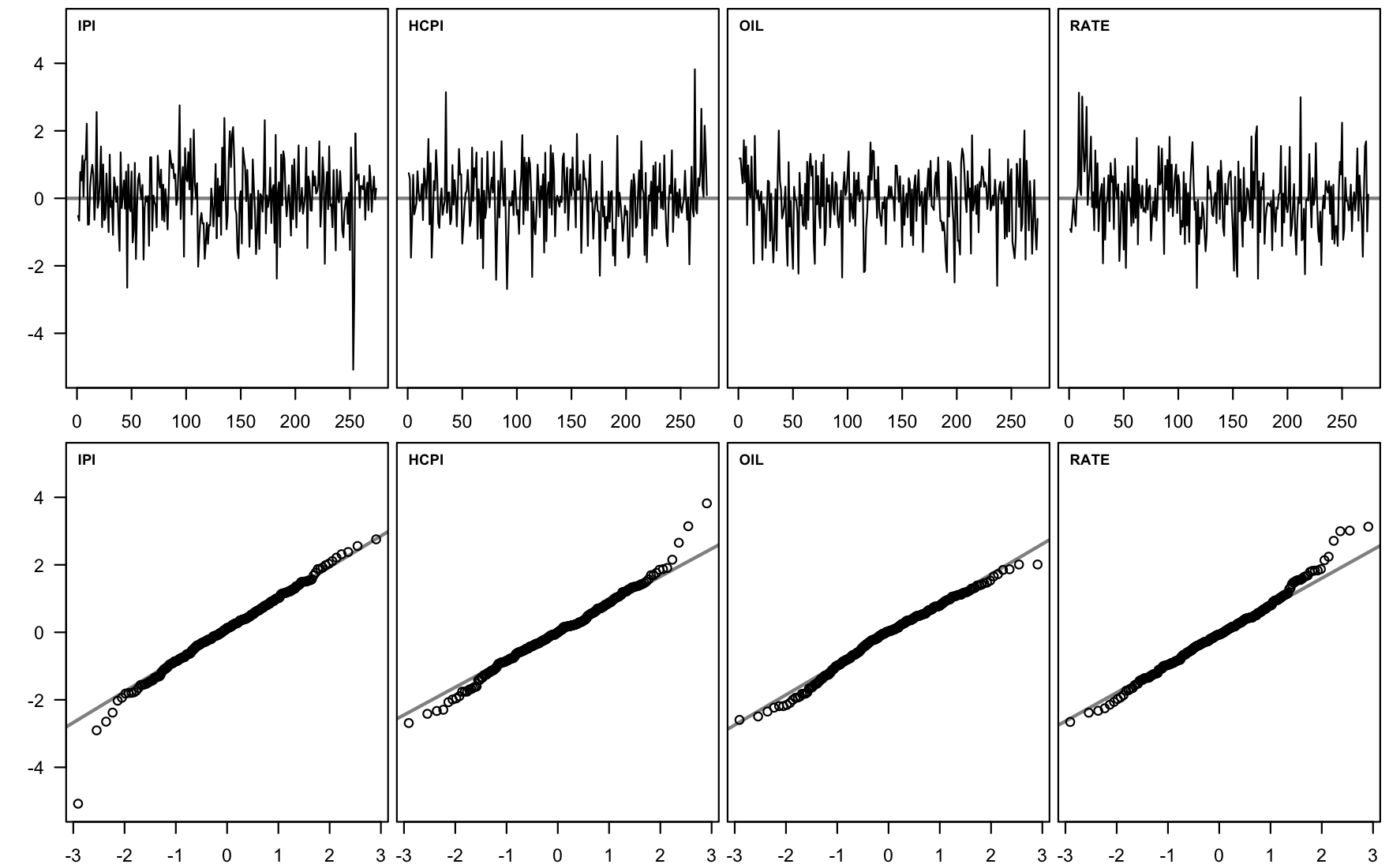}}
    \caption{Quantile residual time series and normal quantile-quantile-plots of the fitted StMVAR($2,2$) model.}
\label{fig:serqqplot}
\end{figure}

The marginal quantile residual time series are presented in the top panels of Figure~\ref{fig:serqqplot}. The time series seem reasonable, as the are no apparent shifts in the mean, volatility, or dynamics. The COVID-19 lockdown shows as a large negative (marginal) quantile residual of IPI, but we do not view this as an inadequacy, as the fast drop in the business cycle is know to be caused by an exceptionally large exogenous shock, and a correctly specified model should thereby produce a large negative residual. Also, the relatively high inflation rates during the COVID-19 crisis show as consecutive positive (marginal) quantile residuals of HCPI. The normal quantile-quantile plots (the bottom panels of Figure~\ref{fig:serqqplot}) show that the marginal distribution of the series appears to be captured relatively well. Overall, we find the adequacy of our model reasonable enough for impulse response analysis.

\subsection{Characteristics of the regimes}\label{sec:characteristics}
Table~\ref{tab:regimes} presents the estimates for the mixing weight parameters, degrees of freedom parameters, and the estimated unconditional means and standard deviations of each marginal series (based on the stationary properties of our model) for both regimes. The mixing weight parameters have the interpretation of being the unconditional probabilities for an observation being generated from each regime. For a correctly specified model, they should hence approximately reflect the proportions of observations generated from each regime. Regime 1 has a mixing weight parameter estimate $0.77$, and it covers approximately $63\%$ of the series (approximated as the mean of the estimated mixing weights), whereas Regime 2 has the implied mixing weight parameter estimate $0.23$ and it covers approximately $37\%$ of the series. Therefore, they are somewhat disproportionate, but seem reasonable enough not to distort the generalized impulse response functions too much. 

\begin{table}[t]
\small
\centering
\begin{tabular}{c c c c c c c c c c c}
               & & & \multicolumn{2}{c}{IPI} & \multicolumn{2}{c}{HCPI} & \multicolumn{2}{c}{OIL} & \multicolumn{2}{c}{RATE} \\
                & $\hat{\alpha}_m$ & $\hat{\nu}_m$ & $\hat{\mu}_{m,1}$ & $\hat{\sigma}_{m,1}$ & $\hat{\mu}_{m,2}$ & $\hat{\sigma}_{m,2}$  & $\hat{\mu}_{m,3}$ & $\hat{\sigma}_{m,3}$  & $\hat{\mu}_{m,4}$ & $\hat{\sigma}_{m,4}$ \\ 
\hline\\[-1.5ex]
Regime 1 & $0.77$ & $\phantom{1}3.40$ & $-2.89$ & $10.55$ & $0.12$ & $0.28$ & $0.02$ & $1.34$ & $0.24$ & $8.89$ \\
Regime 2 & $0.23$ & $12.89$ & $\phantom{-}2.79$ & $\phantom{1}3.20$ & $0.19$ & $0.15$ & $0.33$ & $0.87$ & $2.71$ & $0.69$ \\
\hline
\end{tabular}
\caption{Mixing weight parameter estimates ($\hat{\alpha}_m$), degrees of freedom parameter estimates ($\hat{\nu}_m$), and marginal stationary means ($\hat{\mu}_{m,i}$) and standard deviations ($\hat{\sigma}_{m,i}$) of the component series implied by the fitted StMVAR($2,2$) model for each of the regimes.}
\label{tab:regimes}
\end{table}

The estimates of the degrees of freedom parameter in Table~\ref{tab:regimes} show that Regime 1 has fatter tailed distribution Regime 2. Regime 1 also has negative and volatile long-run output gap, while Regime 2 has positive and less volatile long-run output gap. Long-run inflation is low in Regime 1 (roughly $1.4\%$ yearly), whereas it is moderate in Regime 2 (roughly $2.3\%$ yearly). Also oil price inflation is relatively low in Regime 1, whereas it is high in Regime 2.\footnote{The log-difference of oil price was multiplied by $10$ and not $100$ for numerical reasons, so the unconditional means should be multiplied $10$ to obtain estimates for the (approximate) monthly long-run oil price inflation in percentage units.} The interest rate variable has low mean in Regime 1, but the standard deviation is high, which reflects the wandering movements of the \cite{Wu+Xia:2016} shadow rate after the early 2010's recession. In Regime 2, the interest rate variable has moderate mean and low variance. According to the unconditional variances of the observable variables in Table~\ref{tab:regimes}, Regime 1 is overall more volatile than Regime 2. 

According to the estimated mixing weights presented in Figure~\ref{fig:seriesplot}, Regime 1 mainly prevails after the collapse of Lehman Brothers in the Financial crisis in September 2008. Regime 1 also obtains large mixing weights during and before the early 2000's recession, however. Regime 2 dominates when the first one does not; that is, mainly before the Financial crisis, but excluding the aforementioned periods when the first regime obtains large mixing weights. After the Financial crisis, mixing weights of Regime 2 stay close to zero, excluding a short period before the early 2010's recession.

%

\end{appendices}

\end{document}